\documentclass[aps,prl,superscriptaddress,reprint,floatfix,longbibliography,showpacs,amsfonts,amsmath,amssymb]{revtex4-2}
\usepackage{graphicx}
\usepackage{hyperref}
\usepackage{epstopdf}
\usepackage[note-name=, use-sort-key = false]{notes2bib}
\usepackage{color}
\usepackage{pdfpages}

\hypersetup{
    colorlinks = true,
    citecolor = blue,
    bookmarksnumbered = true,
    urlcolor = blue
}

\makeatletter
\AtBeginDocument{\let\LS@rot\@undefined}
\makeatother

\begin{document}

\title{Near room-temperature ferromagnetism from double-exchange in the van der Waals material CrGeTe$_3$: evidence from optical conductivity under pressure}

\author{Jihaan Ebad-Allah}
\affiliation{Experimentalphysik II, Institute for Physics, University of Augsburg, 86135 Augsburg, Germany}
\affiliation{Department of Physics, Tanta University, 31527 Tanta, Egypt}

\author{Daniel Guterding}
\affiliation{Technische Hochschule Brandenburg, Magdeburger Straße 50, 14770 Brandenburg an der Havel, Germany}

\author{Meera Varma}
\affiliation{Experimentalphysik II, Institute for Physics, University of Augsburg, 86135 Augsburg, Germany}

\author{Mangesh Diware}
\affiliation{Advanced Research Division, Park Systems Co., Suwon 16229, Republic of Korea}

\author{Shraddha Ganorkar}
\affiliation{School of Mechanical Engineering, Sungkyunkwan University, 2066 Seobu-ro, Jangan-gu, Suwon, Gyeonggi-do 16419, Republic of Korea}

\author{Harald O. Jeschke}
\affiliation{Research Institute for Interdisciplinary Science, Okayama University, Okayama 700-8530, Japan}

\author{Christine A. Kuntscher}
\affiliation{Experimentalphysik II, Institute for Physics, University of Augsburg, 86135 Augsburg, Germany}

\begin{abstract}
The unexpected discovery of intrinsic ferromagnetism in layered van der Waals materials has sparked interest in both their fundamental properties and their potential for novel applications. Recent studies suggest near room-temperature ferromagnetism in the pressurized van der Waals crystal CrGeTe$_3$. We perform a comprehensive experimental and theoretical investigation of magnetism and electronic correlations in CrGeTe$_3$, combining broad-frequency reflectivity measurements  with density functional theory and dynamical mean-field theory calculations. Our experimental optical conductivity spectra trace the signatures of developing ferromagnetic order and of the insulator-to-metal transition (IMT) as a function of temperature and hydrostatic pressure. With increasing pressure, we observe the emergence of a mid-infrared feature in the optical conductivity, indicating the development of strong orbital-selective correlations in the high-pressure ferromagnetic phase. We find a distinct relationship between the plasma frequency and Curie temperature of CrGeTe$_3$, which strongly suggests that a double-exchange mechanism is responsible for the observed near room-temperature ferromagnetism. Our results clearly demonstrate the existence of a charge-transfer gap in the metallic phase, ruling out its previously conjectured collapse under pressure.
\end{abstract}

\maketitle

Layered transition-metal chalcogenides host a variety of interesting physical properties and exotic phases such as charge-density wave~\cite{Hwang.2024}, Mott insulator~\cite{Qiao.2017, Kim.2019, Wang.2020b, Tian.2023}, superconductor~\cite{Shi.2015, Qi.2016, Wang.2017}, antiferromagnetic topological insulator~\cite{Otrokov.2019, Li.2019, Li.2019b, Deng.2020}, Weyl semimetal~\cite{Lv.2015, Xu.2015, Lu.2024}, ferromagnetic (FM) nodal-line semimetal~\cite{Kim.2018, Wang.2018, Yun.2023, Wang.2023}, and topological magnon insulator~\cite{Zhu.2021, Luo.2023, Zhuo.2023}, which emerge due to electronic correlations and non-trivial topology in these materials. 
Furthermore, two-dimensional (2D) transition-metal chalcogenides provide an ideal platform for realizing atomically thin van der Waals (vdW) crystals with unique properties such as 2D ferromagnetism~\cite{Gong.2017, Burch.2018} or topological quantum spin Hall effect~\cite{Qian.2014, Cazalilla.2014}, as well as excellent prospects for novel applications~\cite{Deng.2018, Wang.2018b, Gong.2019b, Verzhbitskiy.2020}.

The 2D vdW transition-metal trichalcogenide CrGeTe$_{3}$ and its sibling CrSiTe$_{3}$ are prominent examples for such materials \cite{Zhu.2021, Lin.2016, Liu.2016, Cai.2020}, especially since the discovery of long-range FM order in few-layer CrGeTe$_3$~\cite{Gong.2017}. Each layer of this material forms a honeycomb network of edge-sharing octahedra with a central Cr atom bonded to six Te atoms~\cite{Carteaux.1995, Ji.2013}, as illustrated in the Supplemental Material~\cite{Suppl} (see also references \cite{Bhoi.2021,Watson.2020, Zhang.2019,Xu.2023,Mao.1986, Syassen.2008,Eremets.1992,EbadAllah.2019, EbadAllah.2021, EbadAllah.2022,Tanner.2015,Lu.2017,Uykur.2018,Yang.2021,Koepf.2024,Koepernik.1999,Perdew.1996,Yu.2019,Fang.2018,Shinaoka.2021,Werner.2006, Gull.2011,Otsuki.2024,Aichhorn.2016,jo-cthyb,Werner.2006,Gull.2011,Kim.2020,Fujiwara.2022,Jang.2021,Wang.2020,Leonov.2020,Kang.2022,Kajueter.1996, Parcollet.1999, Kyung.2006, Logan.2016,Katanin.2005,Vidberg.1977,Otsuki.2020,Arsenault.2012,Eschrig.2009,Kang.2021,Pereira.2004,Anderson.1955,Shao.2020,Degiorgi.2011,Qazilbash.2009} therein). At ambient pressure, CrGeTe$_{3}$ is a charge transfer insulator with an energy gap between 0.2 and 0.7 eV~\cite{Ji.2013, Li.2018, Suzuki.2019}. It shows FM order below the Curie temperature $T_{\rm C}=61$-67~K~\cite{Bhoi.2021, Ji.2013, Liu.2017}.

The generally weak vdW forces between layers make these materials susceptible to tuning by application of external pressure, which may lead to structural, magnetic or purely electronic phase transitions~\cite{Cai.2020, Bhoi.2021, Spachmann.2022, Pan.2023, Krottenmuller.2020, Koepf.2024}. Recent transport studies revealed that CrGeTe$_{3}$ is indeed sensitive to external pressure, which leads to a correlated metallic state and near room-temperature ferromagnetism above $\sim3$~GPa~\cite{Bhoi.2021, Xu.2023}. X-ray diffraction and Raman scattering studies show the absence of a structural phase transition in CrGeTe$_{3}$ up to 10 GPa~\cite{Sun.2018}. Furthermore, electron doping via chemical intercalation~\cite{Wang.2019} or electrostatic gating~\cite{Verzhbitskiy.2020}, tensile strain via heterostructuring~\cite{Idzuchi.2023, Zhang.2019c, Dong.2020}, and amorphization due to irradiation~\cite{Zhang.2023} all lead to greatly enhanced Curie temperatures in CrGeTe$_{3}$.

Naturally, this has led to a debate on the mechanism responsible for the enhanced $T_{\rm C}$. While the ferromagnetism of pristine CrGeTe$_{3}$ at ambient pressure has been conclusively attributed to FM superexchange~\cite{Watson.2020, Zhang.2019}, the situation is less clear for the aforementioned modifications. Some authors explain the increase in Curie temperature with the onset of double-exchange interactions upon electron doping~\cite{Verzhbitskiy.2020, Wang.2019}, while another study reports a rise in $T_{\rm C}$ upon hole doping~\cite{Zhang.2023}. Yet another study attributes the rise in $T_{\rm C}$ upon lattice expansion to a weakening of competing antiferromagnetic exchange paths~\cite{Idzuchi.2023}. Under pressure, the increase in Curie temperature has been explained in terms of a decreasing charge transfer gap, which may enhance FM superexchange~\cite{Bhoi.2021}, while we could not definitively pin down the mechanism of enhanced $T_{\rm C}$ in our previous theoretical study~\cite{Xu.2023}. 

\begin{figure*}[t]
\includegraphics[width=1\textwidth]{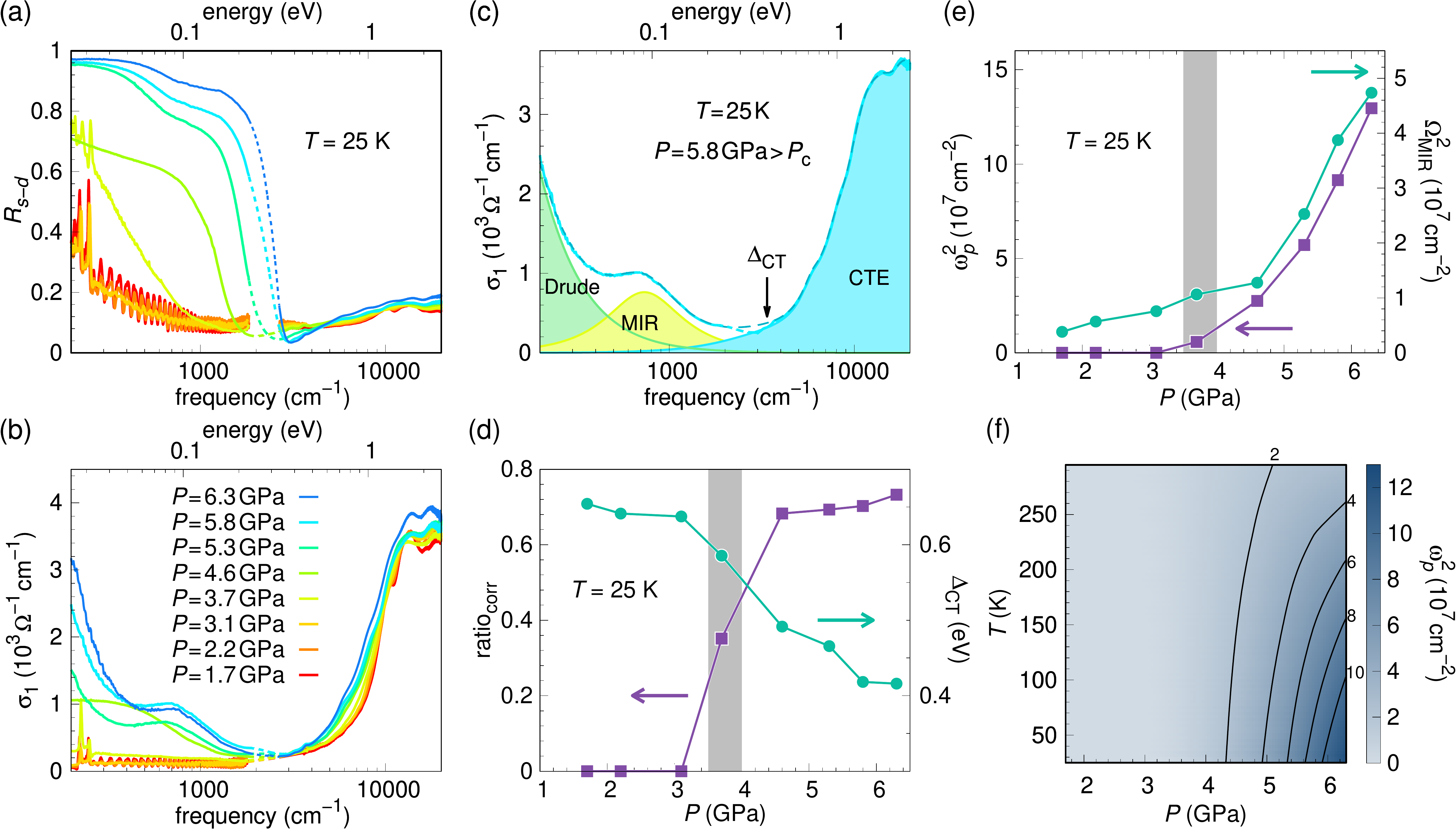}
\caption{
Pressure-dependent (a) reflectivity $R_{s-d}$ and (b) real part of the optical conductivity $\sigma_{1}$ of CrGeTe$_{3}$ at 25 K. (c) Optical conductivity spectrum at 25 K and 5.8 GPa together with the Drude-Lorentz fit and contributions. The charge-transfer gap $\Delta_{\rm CT}$ is indicated by the arrow. (d) Pressure-dependent correlation ratio ratio$_\text{corr}$ (see text for definition) and charge-transfer gap size $\Delta_{\rm CT}$, all at 25 K. (e) Pressure-dependent Drude oscillator strength $\omega^2_p$ and oscillator strength $\Omega^2_\text{MIR}$ of the MIR band, all at 25 K. The vertical gray bar in (d) and (e) indicates the critical pressure $P_c$ of the IMT. (f) Contour map of Drude oscillator strength $\omega^2_p$. The shift of the magnetic transition temperature T$_{C}$ with pressure is indicated by the dashed line.}
\label{fig:MIT}
\end{figure*}

In the present study we attempt to resolve this controversy by analyzing the optical conductivity of the vdW ferromagnet CrGeTe$_3$ under hydrostatic pressure in a combined experimental and theoretical approach. Using broad-frequency reflectivity measurements as well as density functional theory (DFT) and dynamical mean-field theory (DMFT) calculations, we clarify the effect of both magnetism and electronic correlations on the electrodynamic response across the insulator-to-metal transition (IMT). Our optical conductivity measurements show clear signatures of developing FM order as a function of temperature and pressure. In the high-pressure metallic phase, a mid-infrared (MIR) feature emerges in the optical conductivity, which we can attribute to strong orbital-selective electronic correlations. Our data unveils a distinct relationship between the observed plasma frequency $\omega_p$ and the ordering temperature $T_{\rm C}$ of CrGeTe$_3$, which clearly points to a double-exchange mechanism~\cite{Pereira.2004}. Our data show that the optical gap persists into the metallic phase, ruling out a previously conjectured collapse of the charge transfer gap under pressure~\cite{Bhoi.2021}. This, however, does not exclude the relevance of FM superexchange under pressure (see Ref.~\cite{Suppl}). Our study highlights the intertwined nature of magnetism and electronic correlations in CrGeTe$_{3}$.

\begin{figure}[t!]
\includegraphics[width=0.48\textwidth]{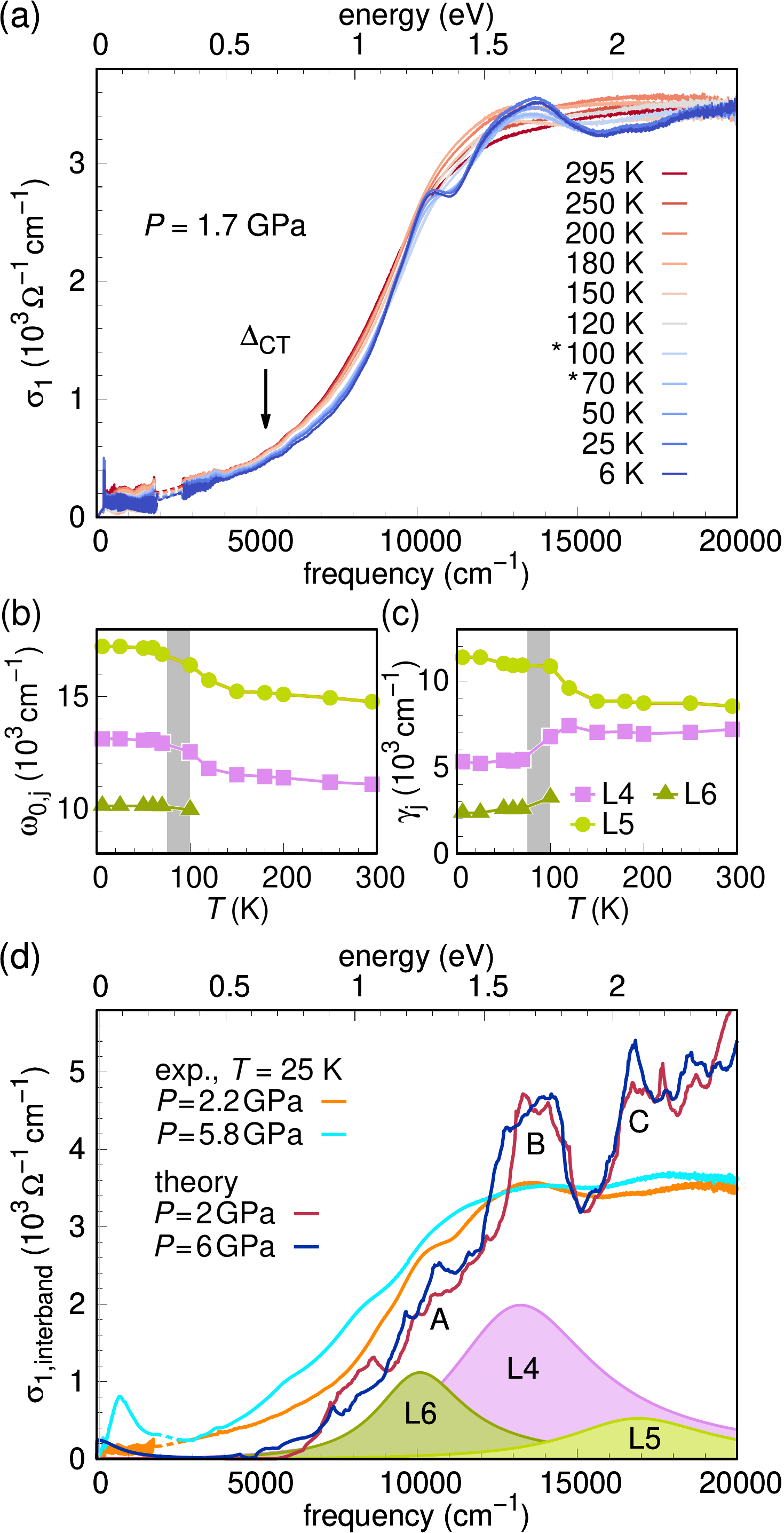}
\caption{
(a) Optical conductivity $\sigma_{1}$ of CrGeTe$_{3}$ at 1.7 GPa at selected temperatures. The magnetic onset temperature is identified by the asterisk. (b) Frequency position and (c) width ($\gamma$) of the L4, L5, and L6 excitations, as obtained from Lorentz fits at 1.6 GPa as a function of temperature. The vertical gray bar in (b) and (c) indicates the Curie temperature T$_C$. (d) Comparison between the experimental and theoretical interband conductivity $\sigma_{1, Interband}$ at 25 K, i.e., within the ferromagnetic phase, together with the fit contributions L4, L5, and L6 at 2.2 GPa.}
\label{fig:spectrumlowpressure}
\end{figure}

We obtain the optical conductivity spectrum throughout the pressure-temperature phase diagram of CrGeTe$_{3}$ by measuring the reflectivity of a small piece of single crystal, loaded into a diamond anvil pressure cell, using Fourier-transform infrared spectroscopy at cryogenic temperatures (see Ref.~\cite{Suppl} for details). The reflectivity $R_{s-d}$ of CrGeTe$_{3}$ at 25 K is displayed in Fig.~\ref{fig:MIT}(a) for selected pressures. Up to 3.1 GPa, the overall reflectivity spectrum changes only modestly. For pressures of 3.7 GPa and above, the low-energy part (below 0.37 eV, \textit{i.e.} 3000 cm$^{-1}$) of $R_{s-d}$ changes drastically: Fabry-Perot interference disappears, the reflectivity level suddenly increases, phonon modes assigned to Te-Cr-Te bending and Cr-Te stretching modes are screened, and a plasma edge develops above 4.6 GPa. Further application of pressure continuously increases the low-energy reflectivity level, while the plasma edge becomes more distinct and shifts to higher energies. Interestingly, a step-like plateau related to an MIR band (see below) develops below the plasma edge. The above observations are evidence of a pressure-induced IMT, consistent with recent reports~\cite{Bhoi.2021,Matsumoto.2024, Xu.2023}. We note here that the critical pressure of the IMT in our study (3.7~GPa) is slightly lower than the one reported in Ref.~\cite{Bhoi.2021} (7 GPa), but in very good agreement with recent electrical transport studies~\cite{Matsumoto.2024, Scharf.2024}.

The signatures of the IMT are also revealed in the real part of the optical conductivity $\sigma_{1}$ (see Fig.~\ref{fig:MIT}(b)), obtained from the $R_{s-d}$ spectra via Kramers-Kronig analysis. For pressures below 3.7 GPa, $\sigma_{1}$ consists of three pronounced high energy excitations at around 2.1, 1.6, and 1.3 eV (L5, L4, and L6 excitations, resp., in Fig.~\ref{fig:spectrumlowpressure}(d)), followed towards lower energies by a drop, marking an absorption edge at $\sim$ 0.65 eV, and two phonon modes at around 0.032 and 0.027 eV. The phonon modes exhibit only minor changes, either by pressure or temperature, and could be assigned to Te-Cr-Te bending and Cr-Te stretching modes, as in CrSiTe$_3$~\cite{Casto.2015}. The absence of the Drude term in the low-pressure regime confirms the ambient-pressure insulating phase consistent with previous reports \cite{Bhoi.2021, Xu.2023, Matsumoto.2024}. From the energy position of the absorption edge, we can also estimate the charge-transfer gap size $\Delta_{\rm CT}$ (see Ref.~\cite{Suppl} for details), depicted in Fig.~\ref{fig:MIT}(d) as a function of pressure at 25 K. Accordingly, the charge-transfer gap is almost constant up to $\sim$3 GPa, whereas above $\sim$3~GPa it starts to decrease in a moderate fashion.

Above the critical pressure $P_c \approx 3.7$ GPa, we need to include a Drude term into the fitting model for the optical response to account for the presence of itinerant charge carriers. With this Drude term, the model captures the sudden increase in the low-energy region of $\sigma_{1}$ (see Fig.~\ref{fig:MIT}(c) as an example), consistent with the recently reported increase in carrier concentration under pressure~\cite{Matsumoto.2024}. Within the metallic regime, the Drude oscillator strength $\omega^2_p$, which is the square of the plasma frequency $\omega_p$, increases with increasing pressure (see Fig.~\ref{fig:MIT}(e)). $\omega^2_p$ serves as a measure to trace the metallic regions in the pressure-temperature phase diagram of CrGeTe$_3$ depicted in Fig.~\ref{fig:MIT}(f).

The presence of itinerant charge carriers leads to a screening of the low-frequency phonon modes. Additionally, an absorption band appears at $\sim 0.09$ eV (MIR band) (see Fig.~\ref{fig:MIT}(c)) and its oscillator strength $\Omega^2_\text{MIR}$ increases with pressure, as shown in Fig.~\ref{fig:MIT}(e). Simultaneously, the charge-transfer gap $\Delta_{\rm CT}$ suddenly shrinks around $P_{\rm c}$ (see Fig.~\ref{fig:MIT}(d)), indicating major changes in the electronic band structure. The charge-transfer gap is slightly reduced from $\sim 0.65$ to 0.4 eV, but does not close even at the highest measured pressure, consistent with our theoretical calculations (see Ref.~\cite{Suppl}). A finite $\Delta_{\rm CT}$ in the metallic state contradicts the assumptions made in Ref.~\cite{Bhoi.2021}, where the onset of metallicity in CrGeTe$_3$ at $\sim 7$ GPa was ascribed to the collapse of the charge-transfer gap boosting the interlayer ferromagnetic superexchange interaction.

\begin{figure}[t]
\includegraphics[width=0.48\textwidth]{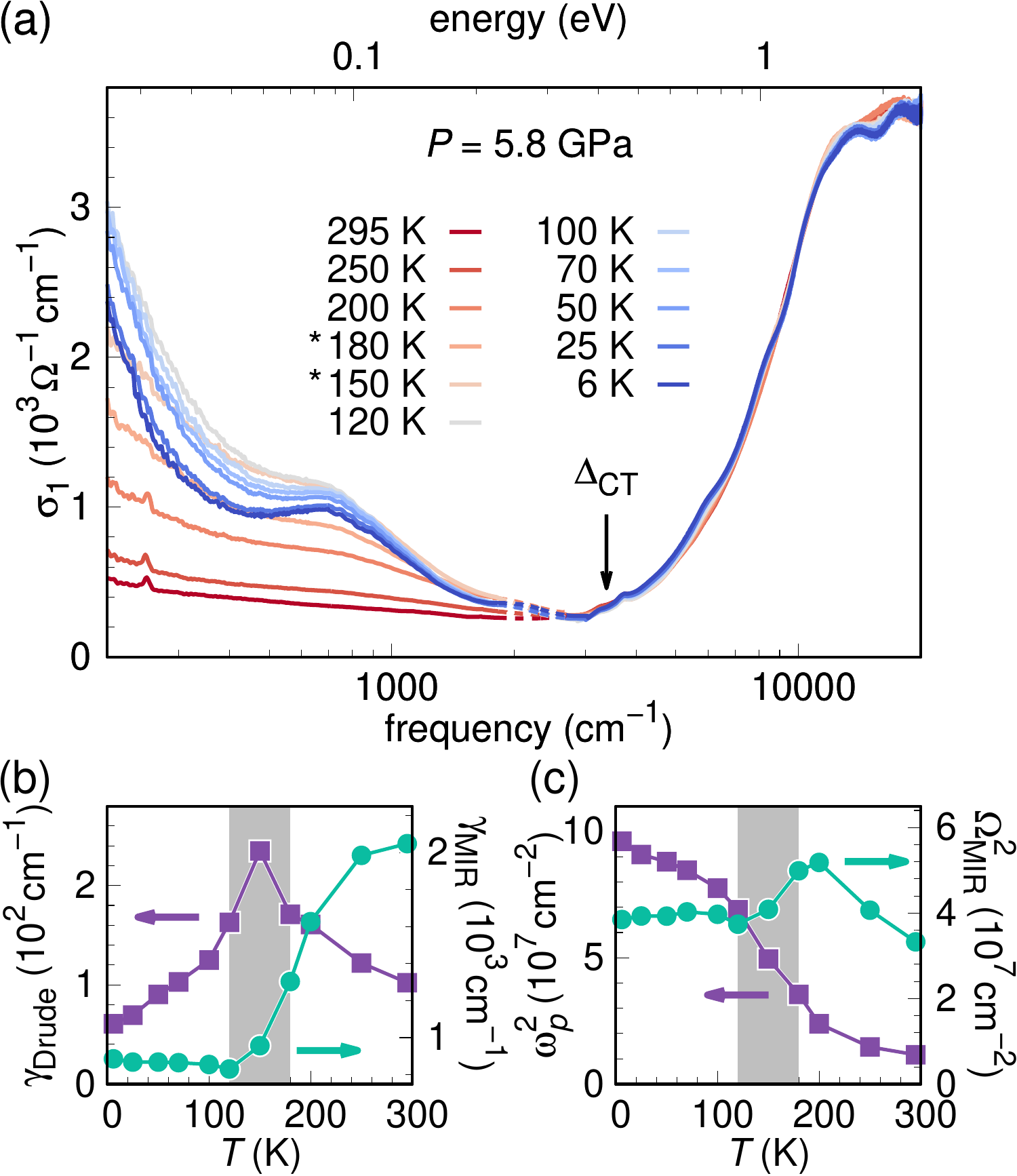}
\caption{
(a) Optical conductivity $\sigma_{1}$ of CrGeTe$_{3}$ at 5.8 GPa at selected temperatures. The magnetic onset temperature is identified by the asterisks. (b) Width and (c) oscillator strength of the Drude term and the MIR oscillator at 5.8 GPa as a function of temperature. The vertical gray bars in (b) and (c) indicate the Curie temperature T$_C$.}
\label{fig:spectrumhighpressure}
\end{figure}

Next, we focus on the effect of pressure on the magnetic ordering in CrGeTe$_{3}$, as revealed by the temperature-induced changes in the optical response at 1.7 GPa (insulating phase, see Fig.~\ref{fig:spectrumlowpressure}(a)) and 5.8 GPa (metallic phase, see Fig~\ref{fig:spectrumhighpressure}(a)). At 1.7 GPa significant changes upon cooling occur in the high-energy excitations (above $\sim 0.7$ eV) in the temperature range 70 - 100~K (see also Ref.~\cite{Suppl}). A decomposition of the optical spectra reveals an anomaly in the frequency position and width ($\gamma$) of the L4 and L5 Lorentz oscillators and the appearance of the L6 excitation during cooling down below 100 K, as displayed by the vertical gray bar in Fig.~\ref{fig:spectrumlowpressure}(b) and (c). The appearance of an additional excitation could also be due to a splitting of the L4 excitation. 

We interpret these excitations in the optical conductivity spectra and their pressure-induced changes using density functional theory (DFT) calculations of electronic band structure and optical conductivity (see Ref.~\cite{Suppl} for details). Fig.~\ref{fig:spectrumlowpressure}(d) displays the pressure evolution of the interband optical conductivity $\sigma_{1, interband}$ in experiment (at 25 K and after subtracting the Drude term) and our DFT calculations. The theoretical calculations agree well with the experimental interband conductivity spectra (see Fig.~\ref{fig:spectrumlowpressure}(d)). In particular, the charge-transfer gap does not close even at the highest studied pressure (6 GPa), which is also supported by the high-pressure electronic band structure (see Ref.~\cite{Suppl}).

The experimental optical conductivity at 25 K contains three prominent features (see Fig.~\ref{fig:spectrumlowpressure}(d)) at around 1.3~eV (A), 1.6~eV (B) and 2.1~eV (C). 
The main three features A, B, and C can be explained based on our DFT calculations. The density of states is dominated by Cr $3d$ and Te $5p$ orbitals. Te and Ge states are present for both the majority and minority spins. The Cr $3d$ orbitals are spin split and mostly occupied for the majority spin, while being mostly unoccupied for the minority spin. This occupation pattern due to ferromagnetic ordering generates the specific three-peak structure of Fig.~\ref{fig:spectrumlowpressure}(d), which is nearly unaffected by pressure. For a more detailed analysis see Ref.~\cite{Suppl}.

At 5.8 GPa (metallic phase), the high-energy excitations are affected similarly when cooling down, though the temperature-induced changes are less pronounced and occur already at a higher temperature of $\sim 150$ K (see Fig.~\ref{fig:spectrumhighpressure}(a) and Ref.~\cite{Suppl}). The effect of magnetic ordering on the low-energy Drude and MIR features upon cooling is, however, remarkably strong: At around 150 K the Drude width shows a maximum and its oscillator strength drastically increases with further cooling. The MIR band sharpens and shows anomalous behavior in its oscillator strength (see Figs.~\ref{fig:spectrumhighpressure}(b) and (c)). Two main conclusions can thus be drawn from our optical data: (i) the FM transition temperature is increased by pressure, consistent with Ref.~\cite{Bhoi.2021}, and (ii) the FM ordering strongly affects the optical response. The latter effect is very rare and, according to our knowledge, has been observed up to now only in FM manganese perovskites~\cite{Park.1996, Sarma.1996, Saitoh.1997} and the FM hexaboride EuB$_6$~\cite{Degiorgi.1997,Kim.2008}.

In our previous DFT+DMFT calculations~\cite{Xu.2023} we observed nearly momentum independent features of the Cr $3d_{z^2}$ spectral function about 200~meV above (minority spin) and below (majority spin) the Fermi level at $P=5$~GPa, which are created by electron-electron correlations (see Ref.~\cite{Suppl}). The MIR feature is explained by an optical transition between the Te $5p$ states at the Fermi level to these Cr $3d_{z^2}$ minority spin states about 200~meV above the Fermi level, and is thus a signature of electronic correlations in the metallic phase. Based on our DFT+DMFT calculations we also extracted orbital-resolved mass enhancements, which are strongly differentiated with respect to the orbital and spin species, with Cr $3d_{z^2}$ electrons being by far the most strongly correlated. Our orbital-averaged experimental estimate of the mass enhancement $1 / \text{ratio}_\text{corr} \approx 1 / 0.7 \approx 1.4$ gives similar results (see Fig.~\ref{fig:MIT}(d) and Ref.~\cite{Suppl}). Accordingly, the metallic phase of FM CrGeTe$_3$ is moderately correlated. In comparison with other vdW materials, the correlation strength is similar to that of transition metal dichalcogenides~\cite{Kim.2020, Fujiwara.2022, Jang.2021}, but significantly weaker than in FM CrI$_3$ monolayers~\cite{Kang.2022} (see Table S1 in Ref.~\cite{Suppl}).

\begin{figure*}[t]
\includegraphics[width=\textwidth]{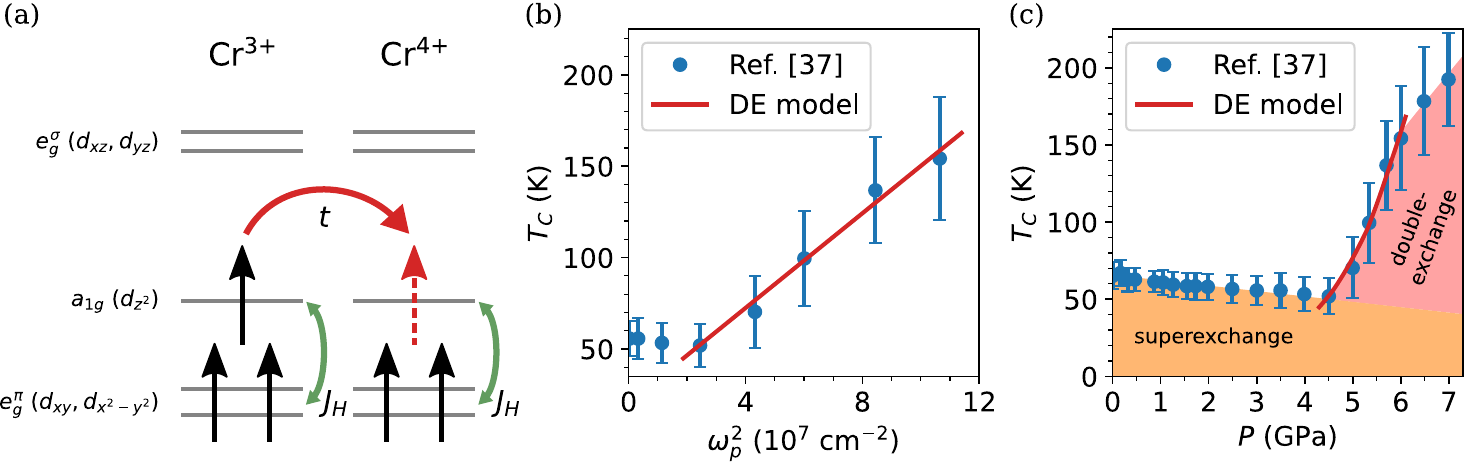}
\caption{(a) Schematic depiction of the double-exchange interaction in CrGeTe$_3$ under high pressure between a $d_{z^2}$ electron on Cr$^{3+}$ and a hole in the $d_{z^2}$ orbital on Cr$^{4+}$. Black arrows denote majority spins. The red dashed arrow represents the majority spin after it has moved into the hole. Horizontal grey lines represent energy levels of Cr $3d$ orbitals. The kinetic term is denoted as $t$ and $J_H$ is the Hund's coupling. (b) Curie temperature $T_{\rm C}$ with experimental error bars from Ref.~\cite{Bhoi.2021} versus the square of the plasma frequency $\omega_p^2$ measured in the present study. The red line represents the linear relationship expected from a double exchange mechanism in the metallic phase~\cite{Pereira.2004}. (c) Curie temperature $T_{\rm C}$ with experimental error bars from Ref.~\cite{Bhoi.2021} versus pressure $P$. The red line represents the Curie temperature $T_{\rm C}$ expected from a double exchange model based on our experimental plasma frequency $\omega_p^2$ as a function of pressure. The double exchange mechanism clearly explains the increase of the Curie temperature when entering the metallic phase. At present, our data only extend to a pressure of up to 6.3 GPa. Therefore, the extension of double-exchange to higher pressures is only a conjecture.}
\label{fig:plasmaCurie}
\end{figure*}

X-ray absorption (XAS) and x-ray magnetic circular dichroism (XMCD) measurements for CrGeTe$_3$ at ambient pressure point to a mixed ground state of Cr$^{3+}$ and Cr$^{2+}$ states due to hybridization between Cr and Te~\cite{Watson.2020}. Consistent with these findings, our previous calculations~\cite{Xu.2023} showed a Cr magnetic moment slightly above $3\,\mu_B$ at ambient pressure, which decreases towards $2.5\,\mu_B$ at around $P=10$~GPa due to the appearance of both holes in the majority spin Cr $3d$ orbitals and electrons in the minority spin Cr $3d$ orbitals. Such a decrease of magnetic moments under pressure is also seen experimentally~\cite{Bhoi.2021}. This suggests that the valence fluctuations of Cr under pressure also include Cr$^{4+}$ states.

While these considerations already suggest that a double-exchange mechanism may be responsible for the sharp increase of the Curie temperature in the metallic phase, our experimental determination of the squared plasma frequency reveals a distinct enhancement in the metallic phase (see Fig.~\ref{fig:MIT}(e)). A mean-field study~\cite{Pereira.2004} of the double-exchange Hamiltonian~\cite{Anderson.1955} shows that the Curie temperature $T_{\rm C}$ should be a linear function of the squared plasma frequency $\omega_p^2$ in double-exchange magnets. In the metallic phase of CrGeTe$_3$, this relation is in excellent agreement with our data and those of Ref.~\cite{Bhoi.2021} (see Fig.~\ref{fig:plasmaCurie}(b)). By matching the values of the plasma frequency back to the pressure axis, we obtain Fig.~\ref{fig:plasmaCurie}(c), which shows that the double-exchange model accurately describes the rise in Curie temperature upon entering the metallic phase. 

Magnetism and correlations in CrGeTe$_3$ are intertwined and strongly differentiated by orbital and spin. Due to the trigonal crystal field, the Cr $3d_{z^2}$ orbital has the highest energy of all occupied orbitals in a Cr$^{3+}$ configuration. Therefore, valence fluctuations in CrGeTe$_3$ mainly affect the Cr $3d_{z^2}$ orbital. In particular, the creation of holes in the majority spin Cr $3d_{z^2}$ orbital enables the double-exchange process, but also affects the behavior of minority spin electrons, which localize around these holes to avoid the Coulomb repulsion of a doubly occupied site (see Ref.~\cite{Suppl}). This localization weakens as more holes are created, which contributes to the observed decrease of correlations (increase of ratio$_\text{corr}$) with further application of pressure in the metallic phase (see Fig.~\ref{fig:MIT}(d)). The minority spin electrons are nevertheless subject to unfavorable Hund's rule interaction (see Ref.~\cite{Suppl}), which we believe leads to the emergence of the MIR feature in the $\sigma_1$ spectrum.

In conclusion, our study shows that the optical conductivity of the van der Waals material CrGeTe$_{3}$ is strongly affected by ferromagnetic ordering and electronic correlations, while being accurately described by DFT+DMFT calculations. The appearance of an MIR feature in the optical conductivity is linked to the emergence of significant electron-electron correlations upon entering the metallic phase, especially in the Cr $3d_{z^2}$ orbital. Based on our measured plasma frequencies, we showed that the increased Curie temperature in the high pressure metallic phase can be explained by a double-exchange mechanism, likely mediated by the pressure-induced creation of holes in the Cr $3d_{z^2}$ orbital. The hereby induced valence fluctuations of Cr could be tested by XAS measurements under pressure. Thus, we can resolve the controversy on exchange interactions in CrGeTe$_3$ under pressure: in the low-pressure insulating regime, FM superexchange dominates, but gradually weakens as pressure causes deviations from the optimal bond geometry; once the system becomes metallic, double-exchange causes the sudden increase in $T_{\rm C}$ to near room-temperature.

According to our microscopic picture of magnetism and electronic correlations in CrGeTe$_3$ both  electron doping and hole doping induces double-exchange mediated near room-temperature ferromagnetism: electron doping via chemical intercalation~\cite{Wang.2019} or electrostatic gating~\cite{Verzhbitskiy.2020} and hole doping by irradiation~\cite{Zhang.2023} or the application of pressure. In the endeavour to design 2D room-temperature magnets, the doping of ferromagnetic van der Waals materials will play an important role, and our study shows how the often invoked double-exchange mechanism can be experimentally substantiated.

\begin{acknowledgments}
H.O.J.~acknowledges fruitful discussions with Han-Xiang Xu and Junya Otsuki. C.A.K.~acknowledges financial support by the Deutsche Forschungsgemeinschaft (DFG), Germany, through Grant No.~KU 1432/15-1. Part of the computation in this work has been done using the facilities of the Supercomputer Center, the Institute for Solid State Physics, the University of Tokyo.
\end{acknowledgments}

\bibliography{referencesCrGeTe3}

\newpage


\section*{Supplementary material}

# Near room-temperature ferromagnetism from double-exchange in van der Waals material CrGeTe₃: evidence from optical conductivity under pressure – Supplemental Material –

Jihaan Ebad-Allah,[1, 2] Daniel Guterding,[3] Meera Varma,[1] Mangesh Diware,[4]
Shraddha Ganorkar,[5] Harald O. Jeschke,[6] and Christine A. Kuntscher[1]

[1] *Experimentalphysik II, Institute for Physics, Augsburg University, 86135 Augsburg, Germany*
[2] *Department of Physics, Tanta University, 31527 Tanta, Egypt*
[3] *Technische Hochschule Brandenburg, Magdeburger Straße 50, 14770 Brandenburg an der Havel, Germany*
[4] *Advanced Research Division, Park Systems Co., Suwon, Republic of Korea*
[5] *School of Mechanical Engineering, Sungkyunkwan University,*
*2066 Seobu-ro, Jangan-gu, Suwon, Gyeonggi-do 16419, Republic of Korea*
[6] *Research Institute for Interdisciplinary Science, Okayama University, Okayama 700-8530, Japan*

## I. SAMPLE PREPARATION, CHARACTERIZATION, AND CRYSTAL STRUCTURE

Single crystals of CrGeTe₃ were grown by using GeTe flux with a mixture of high purity powder of Cr, Ge, and Te at a molar ratio of 2:6:36, as described in detail in Refs. [1, 2]. The rhombohedral crystal structure of CrGeTe₃ with space group $R\bar{3}$ is depicted in Fig. S1. It consists of layers of honeycomb network of edge sharing octahedra formed by central Cr atom bonded to six Te atoms [1, 3] and having a unit order of Te-Ge-Cr-Ge-Te.

Previous studies have explained ferromagnetism in CrGeTe₃ in terms of ferromagnetic superexchange mediated via the close to 90 degrees angle of the Cr-Te-Cr bonds [4, 5]. We show the pressure dependence of this angle in Fig. S2. The angle slightly increases with pressure, which could potentially lead to a weakening of ferromagnetic superexchange, as observed in the pressure range before the insulator-to-metal transition [2]. This weakening is not observed in our previous theoretical study [6], since DFT does not properly capture the paramagnetic insulating state of CrGeTe₃. Ferromagnetic superexchange in a metal may, however, behave more complex than a single leading-order term, as used in Ref. [2].

## II. LOW-TEMPERATURE HIGH-PRESSURE INFRARED REFLECTIVITY MEASUREMENTS

High-pressure reflectance measurements during cooling down from 295 to 6 K were performed for pressures between 1.6 and 6.3 GPa and in the energy range from 0.0248 to 2.48 eV (200 to 20000 cm⁻¹). The measurements were carried out using an infrared microscope (Bruker Hyperion), equipped with a $15x$ Cassegrain objective, coupled to a Bruker Vertex 80v FT-IR spectrometer. A diamond anvil cell (DAC) from EasyLab company equipped with type IIA diamonds, which are suitable for infrared measurements, was utilized for pressure generation. For cooling, a Lake Shore continuous flow cryostat has been used. A freshly cleaved single crystal of CrGeTe₃, with the size of $\sim 160 \times 150 \times 40\ \mu m^3$, was loaded in the hole of a CuBe gasket inside the DAC. For ensuring the well-defined sample-diamond interface throughout the experiment, finely ground CsI powder was used as quasihydrostatic pressure transmitting medium. The pressure was determined *in situ* inside the cryostat using the ruby luminescence method [7, 8]. The pressure-dependent reflectivity spectra at the sample diamond interface $R_{s-d}$ in the energy range 0.0248 to 1.116 eV (200 to 9000 cm⁻¹), were determined according to $R_{s-d}(\omega) = R_{gasket-dia}(\omega) \times (I_s(\omega)/I_{gasket}(\omega))$, where $I_s(\omega)$ is the intensity of the radiation reflected at the interface between the sample and the diamond anvil, $I_{gasket}(\omega)$ the intensity reflected from the CuBe gasket-diamond interface, and $R_{gasket-dia}(\omega)$ is the reflectivity of the gasket material for the diamond interface. While the $R_{s-d}$ spectra in the energy range 1.116 to 2.248 eV (9000–20000 cm⁻¹) were calculated according to $R_{s-d}(\omega) = R_{dia} \times (I_s(\omega)/I_{dia}(\omega))$, where $R_{dia} = 0.167$ is the reflectivity of diamond, which was assumed to be pressure independent [9] and $I_{dia}(\omega)$ is the intensity reflected from the inner diamond-air interface of the empty DAC.

## III. ANALYSIS OF REFLECTIVITY AND OPTICAL CONDUCTIVITY SPECTRA

To obtain the complex optical conductivity $\sigma(\omega) = \sigma_1(\omega) + i\sigma_2(\omega)$, the Kramers-Kronig relations were applied to transform the reflectivity spectra $R_{s-d}$ to the various optical functions, taking the sample-diamond interface into account. The extrapolations of the $R_{s-d}$ spectra were done in a manner similar to our previous publications [10–12]. To this end, Drude-Lorentz fitting procedures were applied for extrapolating the reflectivity data to zero frequency and interpolation in the frequency range 1800-2700 cm⁻¹, which is affected by multiphonon absorptions in the diamond anvils and not completely corrected by the normalization procedure. Above 2.5 eV, we used the high-energy extrapolation of the ambient-pressure reflectivity spectrum obtained by x-ray atomic scattering functions crystals [13] after adjustment for the

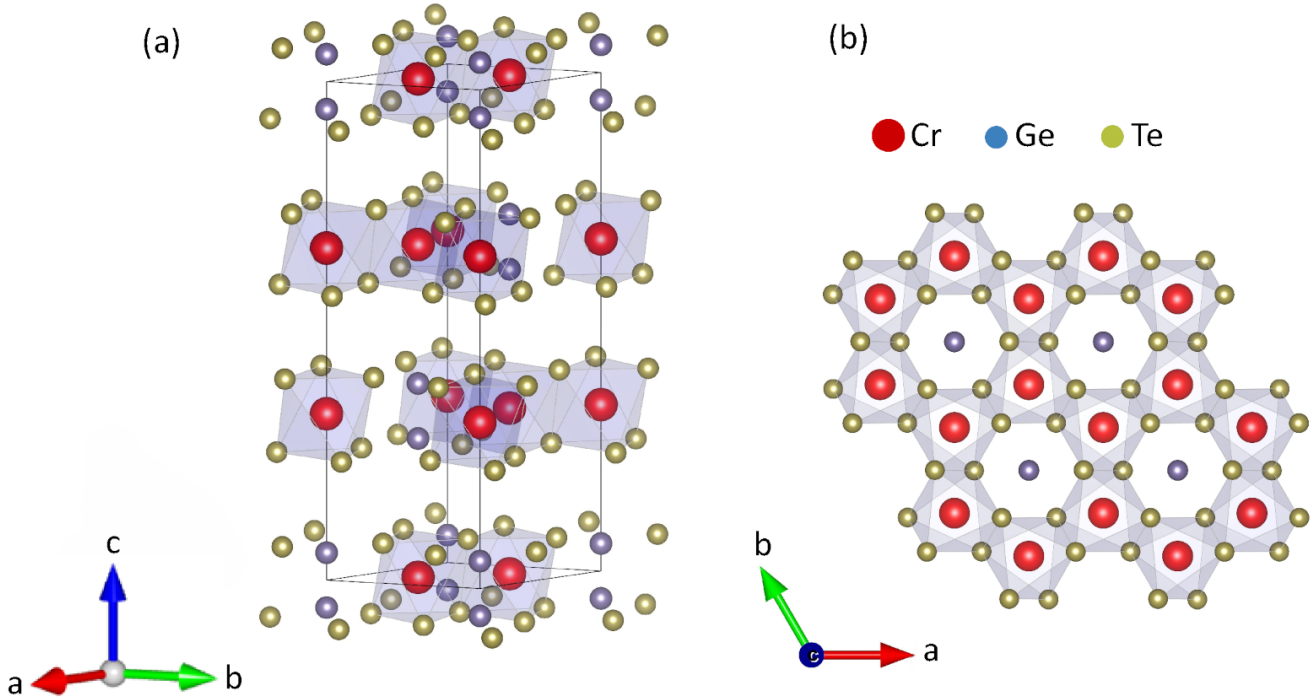

FIG. S1. Rhombohedral crystal structure of CrGeTe$_3$ with space group $R\bar{3}$ consisting of honeycomb layers with edge sharing octahedra formed by central Cr atom bonded to six Te atoms [1, 3].

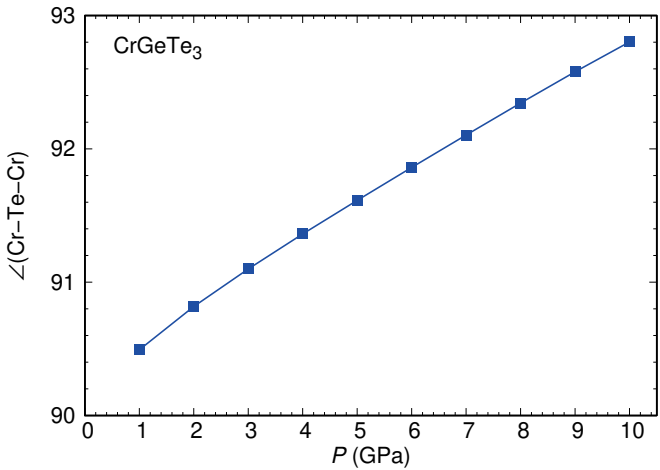

FIG. S2. Pressure evolution of the Cr-Te-Cr angle which corresponds to the nearest neighbour exchange path $J_1$. Structures were interpolated as in Ref. [6].

sample-diamond interface. Furthermore, $R_{s-d}$ and the optical conductivity were simultaneously fitted with the Drude-Lorentz model for decomposition of the $\sigma_1$ spectrum.

Within the Drude-Lorentz model the complex dielectric function $\epsilon(\omega) = \epsilon_1(\omega) + i\epsilon_2(\omega)$ is given as

$$\epsilon(\omega) = \epsilon_\infty - \frac{\omega_{p,Drude}^2}{\omega^2 + i\omega/\tau_{Drude}} + \sum_j \frac{\Omega_j^2}{\omega_{0,j}^2 - \omega^2 - i\omega\gamma_j} \quad, \tag{1}$$

where $\epsilon_\infty$ is the high-energy contribution to $\epsilon_1$. $\omega_{p,Drude}$ and $1/\tau_{Drude}$ are the plasma frequency and scattering rate of itinerant charge carriers, respectively. $\omega_{0,j}$, $\Omega_j$, and $\gamma_j$ are the eigenfrequency, oscillator strength, and width of the j$^{th}$ Lorentz oscillator, respectively. The decomposition of the $\sigma_1$ spectra into Drude and Lorentz contributions as a function of pressure and temperature is given in section VIII.

The optical gap size $\Delta$ was estimated by a linear extrapolation of the absorption edge in the $\sigma_1$ spectrum, as illustrated in Fig. S3 for the conductivity spectrum at 25 K and 5.8 GPa.

## IV. DENSITY FUNCTIONAL THEORY CALCULATIONS

For the interpretation of the observed excitations in the optical conductivity spectra, we perform density functional theory (DFT) calculations within the full potential local orbital (FPLO) method [14] in generalized gradient approximation (GGA) [15] for the exchange-correlation functional. We used experimental crystal structures for CrGeTe$_3$ under pressure from Ref. [16], which we interpo-

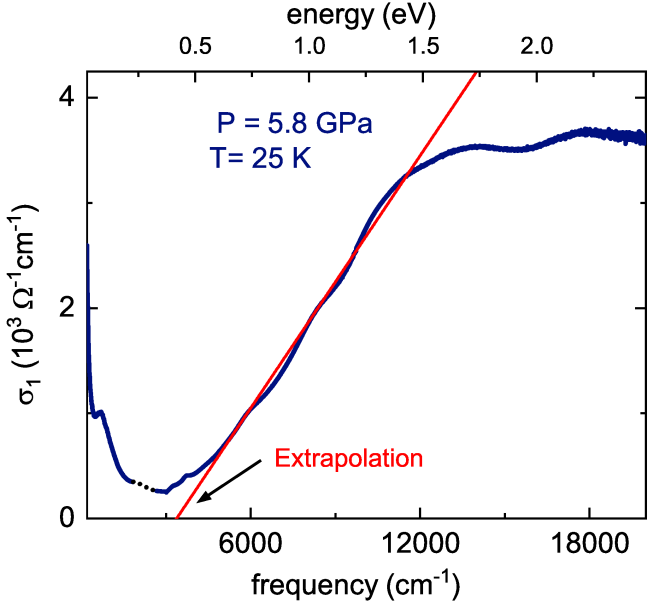

FIG. S3. Linear extrapolation of the absorption edge in the $\sigma_1$ spectrum at 25 K and 5.8 GPa as an example, for determining the optical gap size $\Delta$.

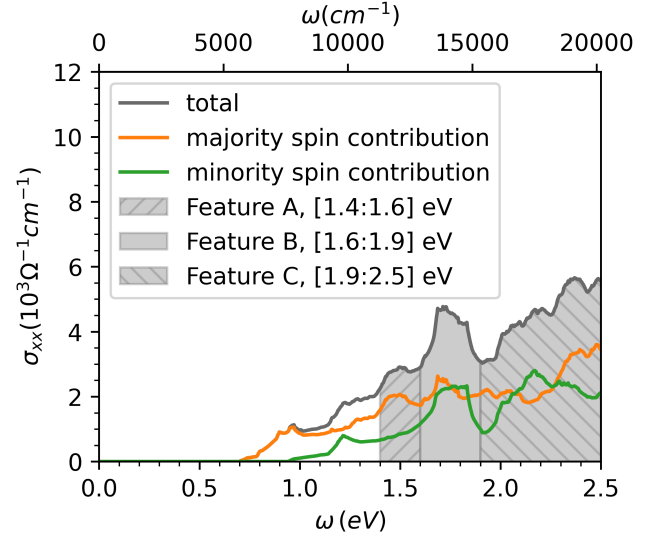

FIG. S5. Spin contributions to the $\sigma_{xx}$ component of the optical conductivity of CrGeTe$_3$ at P = 0 GPa in the ferromagnetic state calculated using DFT. We relate the shaded energy windows to the observed features A, B and C of the measured optical conductivity.

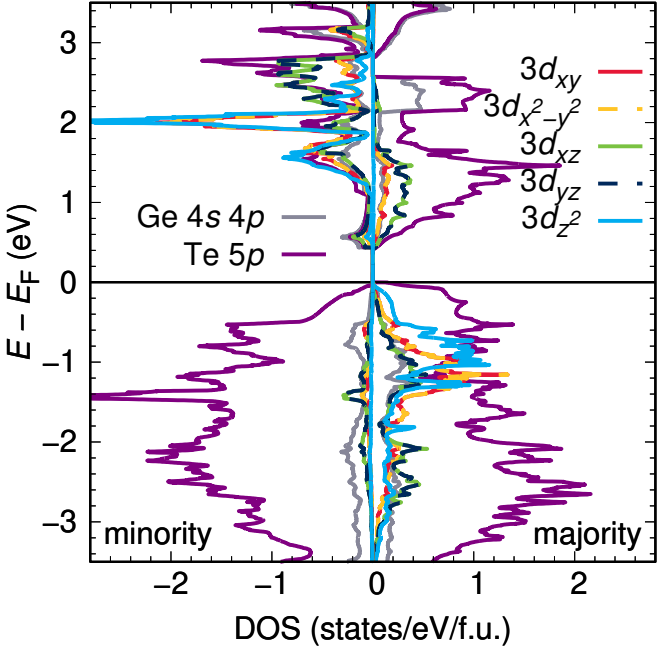

FIG. S4. Orbital-resolved density of states of CrGeTe$_3$ at ambient pressure (P = 0 GPa) in the ferromagnetic state calculated using DFT.

lated smoothly as explained in Ref. [6]. All calculations were performed in ferromagnetic spin configuration.

We calculated the partial density of states (DOS) in the ferromagnetic state at ambient pressure (P = 0 GPa) from FPLO using a $50 \times 50 \times 50$ $k$-point grid. The results are shown in Fig. S4. The density of states is dominated by Cr $3d$ and Te $5p$ orbitals, with additional contributions from Ge $4s$ and $4p$ states. Te and Ge states are present for both majority and minority spins. The Cr $3d$ orbitals are spin split and mostly occupied for the majority spin, while being mostly unoccupied for the minority spin.

The optical conductivity in the main text was calculated using the FPLO density functional theory code on a $50 \times 50 \times 50$ $k$-point grid. For the figures in the supplemental material, we modified the FPLO code to obtain the inter-band contribution to the symmetric (in band indices) band-resolved optical conductivity tensor $[\sigma_{xx}^\pm(\omega)]_{ij}$, where $i$ and $j$ are band indices, which run over all bands in the FPLO basis, and $+/-$ stands for the majority/minority spin. The band-resolved optical conductivity tensor was calculated on a coarser $20 \times 20 \times 20$ $k$-point grid (for performance reasons). We verified that a sum over the band indices of this tensor reproduces the result of the unmodified code. Alternatively, we could have calculated the orbital-resolved optical conductivity tensor, which would have required extensive changes to the FPLO DFT code and caused further numerical effort. Therefore, we chose not to pursue this route.

We calculated the spin-resolved optical conductivity $\sigma_{xx}^\pm(\omega)$ at ambient pressure (P = 0 GPa) using the FPLO method, with $ab\text{-}initio$ inter-band transition matrix elements. The total optical conductivity is defined as the sum of the majority and minority spin component: $\sigma_{xx}(\omega) = \sigma_{xx}^+(\omega) + \sigma_{xx}^-(\omega)$. The results are shown in Fig. S5. We shaded the energy regions, in which we observe three distinct features that are similar to the experimentally observed optical conductivity. These are feature $A$ in the energy region $[1.4 : 1.6]$ eV, feature $B$

in the energy region [1.6 : 1.9] eV and feature $C$ in the energy region [1.9 : 2.5] eV. Under pressure these features are shifted very slightly, but the qualitative picture remains unchanged. The calculated optical conductivity is in good agreement with experimental data.

The experimental optical conductivity at 25 K contains three prominent features (see Fig. 2(d) in main text) at around 1.3 eV (A), 1.6 eV (B) and 2.1 eV (C). Under pressure an additional absorption band arises below 200 meV. The main three features A, B, and C can be explained based on our DFT calculations (see Fig. S5). Feature A is dominated by transitions between Cr $3d$ and Te $5p$ states with majority spin. Features B and C are explained by transitions between Cr $3d$ and Te $5p$ states with both majority and minority spin. The dip in the calculated optical conductivity between features B and C follows from a gap between minority spin Cr $3d$ states above the Fermi level. These transitions are indicated in the schematic DOS in Fig. S6. A previous DFT study of ambient pressure CrGeTe$_3$ predicted similar features in the optical conductivity, but at energies above 2 eV. [17].

In the following we visualize the contributions of each band to the optical conductivity in the specific energy regions. In addition to the standard band weights $b_{ij}^{\pm}(\vec{k})$, which represent the weight of orbital $j$ to band $i$ at a $k$-point $\vec{k}$, we calculate multiplicative band weights $c_i^{\pm}(\omega_{\min}, \omega_{\max}) \in [0,1]$ for each band $i$ and energy interval $\omega \in [\omega_{\min}, \omega_{\max}]$ of the optical conductivity. For this, we integrate over the frequency parameter of the optical conductivity tensor:

$$[\sigma_{xx}^{\pm}]_{ij}(\omega_{\min}, \omega_{\max}) = \int_{\omega_{\min}}^{\omega_{\max}} d\omega \, [\sigma_{xx}^{\pm}(\omega)]_{ij} \qquad (2)$$

The raw weight of each band is then given by the sum over one of the band indices of the tensor:

$$\tilde{c}_i^{\pm}(\omega_{\min}, \omega_{\max}) = \sum_j [\sigma_{xx}^{\pm}]_{ij}(\omega_{\min}, \omega_{\max})$$
$$= \sum_j \int_{\omega_{\min}}^{\omega_{\max}} d\omega \, [\sigma_{xx}^{\pm}(\omega)]_{ij} \qquad (3)$$

For convenience of visualization, we normalize these raw weights so that they faithfully represent the relative contribution of each band to the optical conductivity in the energy region of interest:

$$c_i^{\pm}(\omega_{\min}, \omega_{\max}) = \frac{\tilde{c}_i^{\pm}(\omega_{\min}, \omega_{\max})}{\max\left(\tilde{c}_i^+(\omega_{\min}, \omega_{\max}), \tilde{c}_i^-(\omega_{\min}, \omega_{\max})\right)} \qquad (4)$$

The band weights $a_{ij}^{\pm}(\vec{k}, \omega_{\min}, \omega_{\max})$ we visualize in the following are given by the product of the standard orbital weight $b_{ij}^{\pm}(\vec{k})$ and our custom weight $c_i^{\pm}(\omega_{\min}, \omega_{\max})$:

$$a_{ij}^{\pm}(\vec{k}, \omega_{\min}, \omega_{\max}) = b_{ij}^{\pm}(\vec{k}) \cdot c_i^{\pm}(\omega_{\min}, \omega_{\max}) \qquad (5)$$

Here, $i$ is the band index, while $j$ is the orbital index. Since the weights $c_i^{\pm}(\omega_{\min}, \omega_{\max})$ are zero for any bands, which do not contribute to the optical conductivity in the energy window $[\omega_{\min}, \omega_{\max}]$, this multiplication filters out bands, which are irrelevant for the optical conductivity in the respective energy window. Important bands will be represented by weights proportional to their contribution to the optical conductivity as defined above. Our results for the three energy windows identified in Fig. S5 are shown in Figs. S7, S8 and S9.

Our analysis clearly differentiates the contributions of majority and minority spin electrons and also allows us to analyze the region of active bands for the optical conductivity, as well as the contribution of each orbital to these bands. In these figures we only show orbital weights for Cr $3d$ and Te $5p$ orbitals. Ge $4s$ and $4p$ and any other weights on the relevant bands are relatively small, as can be seen by comparing the band structure to the orbital-resolved density of states (see Fig. S4 and also Fig. S6).

As explained in the main text, the calculated optical conductivity does not change dramatically with pressure, even though the material becomes metallic. At ambient pressure CrGeTe$_3$ has an indirect band gap. Under pressure, the system becomes metallic, but the band gap closes only indirectly (see Fig. S10). Since optical transitions do not transfer momentum, i.e. they occur vertically in our electronic band structure diagrams, the inter-band contribution to the optical conductivity remains zero at low excitation energies due to the vertical gap between highest occupied and lowest unoccupied band at each $k$-point.

## V. DENSITY FUNCTIONAL THEORY + DYNAMICAL MEAN-FIELD THEORY CALCULATIONS

We performed DFT + dynamical mean-field theory (DFT+DMFT) calculations within DCORE [18] for the electronic structure of CrGeTe$_3$, as explained in Ref. [6]. In particular, we used the hybridization expansion continuous-time quantum Monte Carlo (CT-QMC) method to solve the DMFT impurity problem [19, 20].

Our previous analysis of the DFT+DMFT spectral function in the ferromagnetic state at T = 100 K and P = 5 GPa (see Fig. 5d and 5e in Ref. [6]) shows an almost flat in momentum space feature at around +200 meV for the minority spin electrons. A corresponding feature appears in the majority spin spectral function around -200 meV (see Fig. 5e and 5f in Ref. [6]).

We performed additional analysis of the electronic self-energy of CrGeTe$_3$ to explain these features of the spectral function. We can clearly identify the peaks at ±200 meV in the DFT+DMFT spectral function with peaks in the DFT+DMFT electronic self-energy for the Cr $3d_{z^2}$ orbital (Fig. S11). The imaginary part of the self-energy resembles a doped Mott-insulator [27–30], although the energy difference between the two features

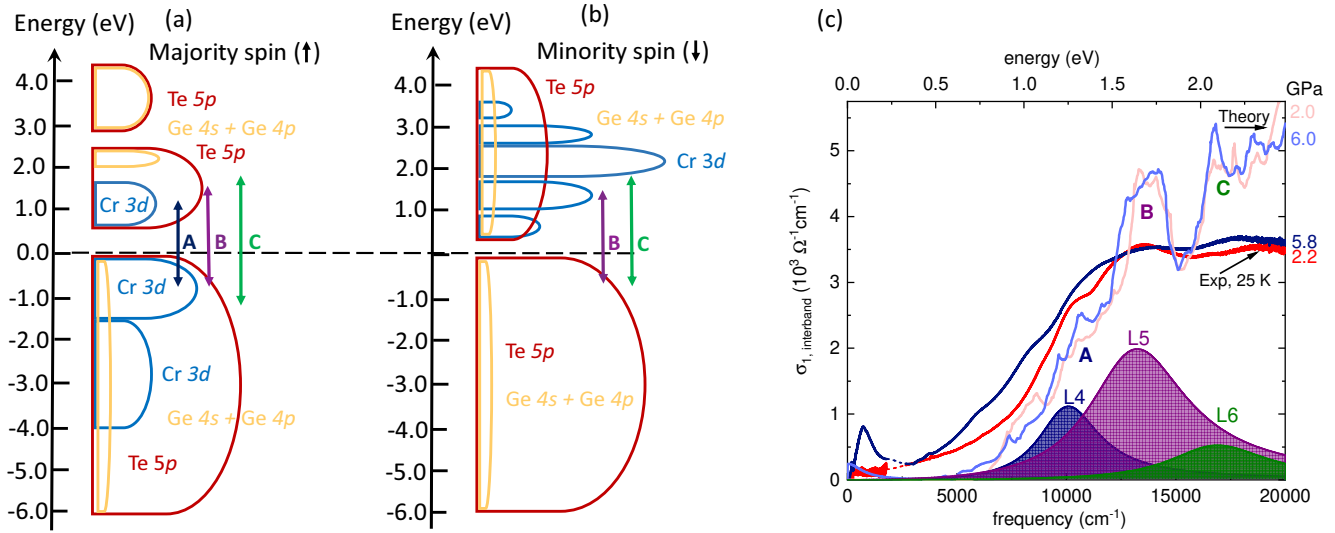

FIG. S6. Energy scheme of CrGeTe$_3$ for (a) majority and (b) minority spin. The vertical arrows mark the possible electronic transitions, which explain the three interband excitations A, B and C. (c) Comparison between the experimental and theoretical interband conductivity $\sigma_{1,Interband}$ at 2.2 and 5.8 GPa and at 25 K, i.e., within the ferromagnetic phase, together with the fit contributions L4, L5, and L6 at 2.2 GPa.

TABLE S1. DFT+DMFT quasiparticle weights for CrGeTe$_3$ and other compounds. In case two values are given for the quasiparticle weight, these represent the values for the majority/minority spin states. For references containing temperature-dependent values for the quasiparticle weight, we take the value corresponding to the lowest available temperature. For orbitals not listed in our table, the cited references did not contain any values for the quasiparticle weight.

| compound | CrGeTe$_3$ ($P = 5$ GPa) [6] | VSe$_2$ [21] | CoS$_2$ [22] | NiSe$_2$ [23] |
|---|---|---|---|---|
| weakly correlated orbitals | Cr 3$d_{xy}$, $d_{x^2-y^2}$, $d_{xz}$, $d_{yz}$ | V 3$d_{xz}$, $d_{yz}$ | Co 3$d_{xy}$, $d_{xz}$, $d_{yz}$ | Ni 3$d_{xy}$, $d_{xz}$, $d_{yz}$ |
| quasiparticle weight | $\sim 0.8$ | $\sim 0.65$ | 1.0 | 1.0 |
| strongly correlated orbitals | Cr 3$d_{z^2}$ | V 3$d_{z^2}$, $d_{xy}$, $d_{x^2-y^2}$ | Co 3$d_{z^2}$, $d_{x^2-y^2}$ | Ni 3$d_{z^2}$, $d_{x^2-y^2}$ |
| quasiparticle weight | $\sim 0.6 / \sim 0.45$ | $\sim 0.4$ | 0.83 / 0.59 | $\sim 0.5$ |
| compound | SrNiO$_2$ [24] | LaNiO$_2$ [24] | NdNiO$_2$ [25] | CrI$_3$ monolayer [26] |
| weakly correlated orbitals | Ni 3$d_{z^2}$, $d_{xy}$, $d_{xz}$, $d_{yz}$ | Ni 3$d_{z^2}$, $d_{xy}$, $d_{xz}$, $d_{yz}$ | Ni 3$d_{z^2}$ | Cr 3$d_{xy}$, $d_{xz}$, $d_{yz}$ |
| quasiparticle weight | $\sim 0.66$ | $\sim 0.81$ | $\sim 0.77$ | $\sim 0.5 / \sim 0.42$ |
| strongly correlated orbitals | Ni 3$d_{x^2-y^2}$ | Ni 3$d_{x^2-y^2}$ | Ni 3$d_{x^2-y^2}$ | Cr 3$d_{z^2}$, $d_{x^2-y^2}$ |
| quasiparticle weight | 0.53 | 0.36 | $\sim 0.34$ | $\sim 0.58 / \sim 0.25$ |

suggests that these are not Hubbard bands. The spin-splitting is induced by the ferromagnetism and resembles previous theoretical results [31].

Therefore, the peaks in the spectral function at $\pm 200$ meV can be interpreted as correlation-induced features of the Cr $3d_{z^2}$ orbital. We conjecture that the formation of these features in the spectral function is connected to the double-exchange mechanism (see below). Certainly, we can identify the mid-infrared (MIR) feature of the optical conductivity as a transition of minority-spin electrons from below the Fermi level to this correlation-induced peak of the Cr $3d_{z^2}$ spectral function above the Fermi level.

The discussion above applies to a pressure of P = 5 GPa. At lower pressures, CrGeTe$_3$ is an insulator.

Pressure induces a transition from an insulator to a correlated ferromagnetic metal due to the creation of holes in the majority-spin states and electrons in the minority-spin states of chromium, as well as an overall increase in Cr $3d$ occupancy (see Fig. 12 in Ref. [6]). Therefore, we believe that the inherent occupation imbalance of majority and minority spin Cr $3d$ states under pressure helps to avoid an insulating state.

We verified that the feature in the self-energy at an energy of about +200 meV is not an artifact of the analytic continuation procedure, which is employed when working with CT-QMC impurity solvers. In Ref. [6] we used the Padé method for analytic continuation [32], which is known to capture well at least the features close to the Fermi level. Here, we additionally used the recently

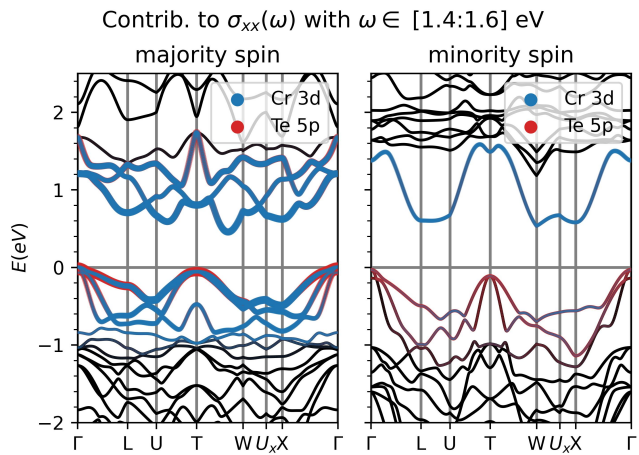

FIG. S7. DFT-calculated spin-resolved electronic band structure of CrGeTe₃ at P = 0 GPa in the ferromagnetic state with orbital weights multiplied by the relative contributions of each band to the optical conductivity in the energy window [1.4:1.6] eV (feature A).

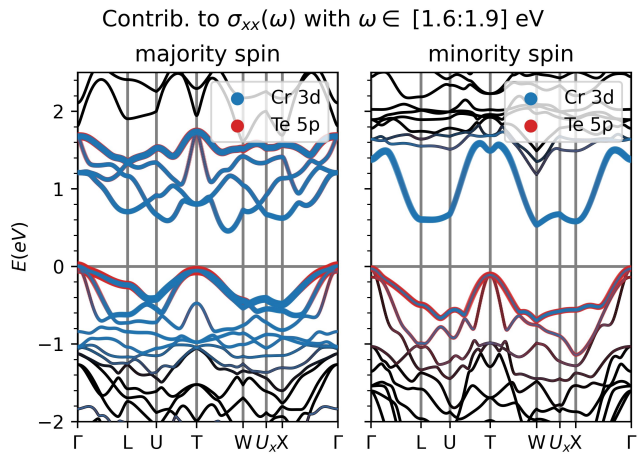

FIG. S8. DFT-calculated spin-resolved electronic band structure of CrGeTe₃ at P = 0 GPa in the ferromagnetic state with orbital weights multiplied by the relative contributions of each band to the optical conductivity in the energy window [1.6:1.9] eV (feature B).

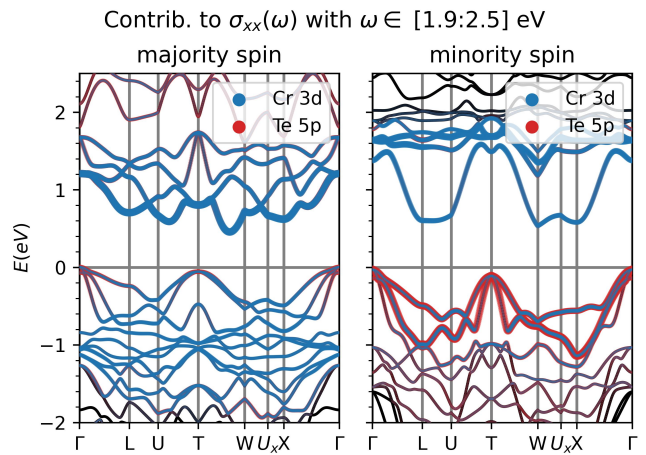

FIG. S9. DFT-calculated spin-resolved electronic band structure of CrGeTe₃ at P = 0 GPa in the ferromagnetic state with orbital weights multiplied by the relative contributions of each band to the optical conductivity in the energy window [1.9:2.5] eV (feature C).

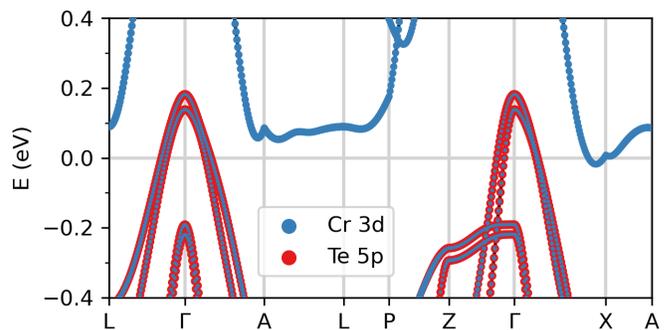

FIG. S10. Full-relativistic electronic band structure of CrGeTe₃ in the ferromagnetic state calculated from DFT at a pressure of 5 GPa. The chosen $k$-path differs from other band structure plots, so that both hole and electron pockets are easily recognizable.

developed sparse modeling (SpM) analytic continuation method [33], which may improve the accuracy at higher energies.

As expected, the electronic self-energy of the minority spin Cr $3d_{z^2}$ orbital is similar in both methods (see Fig. S12). While the Padé result shows only one major feature in both the real and imaginary part of the self-energy, the SpM result contains additional minor features.

Therefore, the spectral function of the minority spin Cr $3d_{z^2}$ orbital is very similar in both methods (see Fig. S13). As expected, the low-energy region of the spectral function is almost identical. The hump in the

$k$-integrated spectral function around +200 meV, which we believe is observed in the experimental optical conductivity under pressure, is present irrespective of the analytic continuation method. Minor differences only appear at energies higher than about +0.5 eV, which the Padé method often does not capture in all detail. This does not affect any of the conclusions of our previous calculations for CrGeTe₃ (see Ref. [6]).

To estimate the correlation strength in CrGeTe₃ at 5 GPa, we calculate in our DFT+DMFT calculations from the electronic self-energy $\Sigma_\sigma^m$ at the lowest positive Matsubara frequency $\omega_0$ the quasiparticle-weight $z_\sigma^m$, where $\sigma$ denotes the spin and $m$ denotes the orbital index [34]. The quasiparticle weight is also the inverse of the mass enhancement over a pure DFT calculation due to correlations, *i.e.* the effective mass $m^*$ divided by the

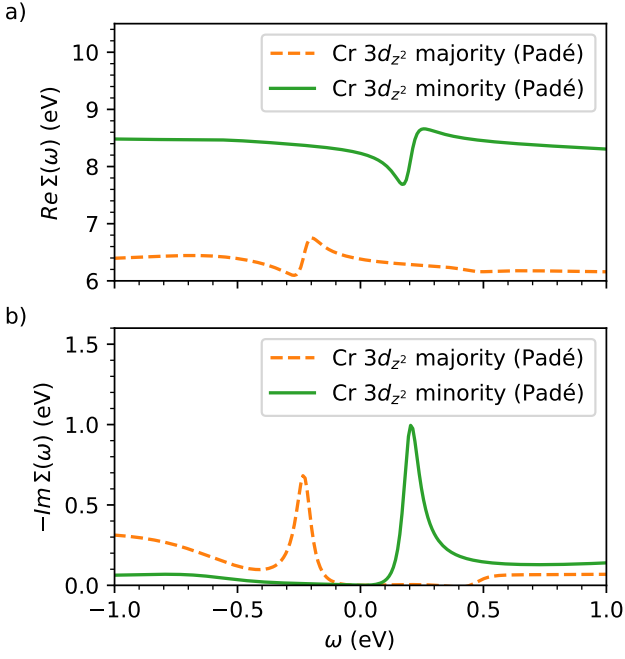

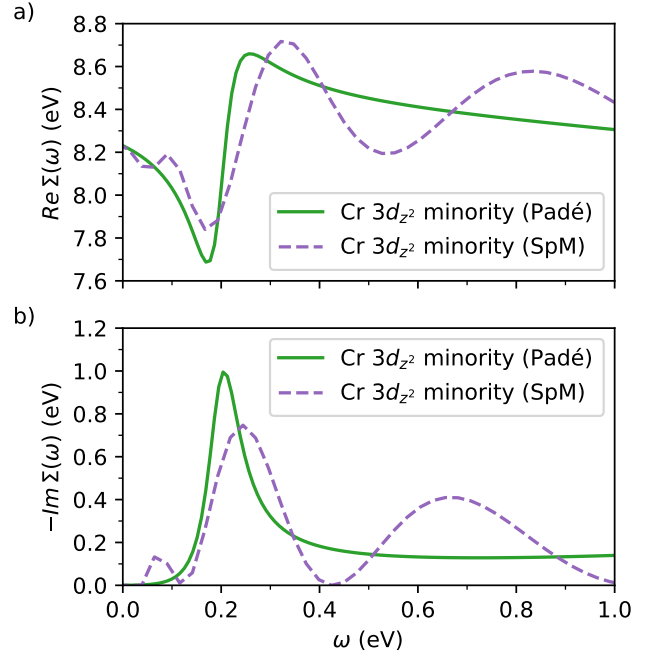

FIG. S11. DFT+DMFT self-energy on the real frequency axis $\Sigma(\omega)$ for the majority- and minority-spin Cr $3d_{z^2}$ orbital of CrGeTe$_3$ at T = 100 K and P = 5 GPa in the ferromagnetic state. Results are shown for the Padé analytic continuation method. a) shows the real part of the self-energy $\mathrm{Re}\,\Sigma(\omega)$. b) shows the imaginary part of the self-energy $-\mathrm{Im}\,\Sigma(\omega)$.

FIG. S12. DFT+DMFT self-energy on the real frequency axis $\Sigma(\omega)$ for the minority-spin Cr $3d_{z^2}$ orbital of CrGeTe$_3$ at T = 100 K and P = 5 GPa in the ferromagnetic state. Results are shown for both Padé and sparse modeling (SpM) analytic continuation methods. a) shows the real part of the self-energy $\mathrm{Re}\,\Sigma(\omega)$. b) shows the imaginary part of the self-energy $-\mathrm{Im}\,\Sigma(\omega)$.

DFT band mass $m_{\mathrm{DFT}}$:

$$(z_\sigma^m)^{-1} = 1 - \frac{\mathrm{Im}\,\Sigma_\sigma^m(\omega_0)}{\omega_0} = \frac{m^*}{m_{\mathrm{DFT}}} \quad . \quad (6)$$

Effects of electronic correlations among the Cr $3d$ orbitals in CrGeTe$_3$ are strongly orbital-selective [6]. The minority spin $a_{1g}$ ($d_{z^2}$) electrons are strongly correlated under pressure, as demonstrated by a decreased quasiparticle weight of about 0.45, which leads to a substantial mass enhancement. The majority spin $a_{1g}$ orbital is slightly less correlated, with a quasiparticle weight of about 0.6. The $e_g^\pi$ ($d_{xy}$, $d_{x^2-y^2}$) and $e_g^\sigma$ ($d_{xz}$, $d_{yz}$) orbitals are weakly correlated with a quasiparticle weight of about 0.8. The strength of electronic correlations in CrGeTe$_3$ under pressure, as measured by the quasiparticle weight, is similar to theoretical results for transition metal dichalcogenides [21–23] and nickelates [24, 25, 35], while monolayers of transition metal trihalides appear to be more strongly correlated [26] (see Table S1).

We note here that the double-exchange picture explains the strong differentiation of effective masses by orbitals and spins in CrGeTe$_3$. Since the Cr $3d_{z^2}$ orbital is closest to half filling and fully polarized, its electrons are most impeded by electron-electron correlations. The creation of holes in the majority spin Cr $3d_{z^2}$ orbital not only mobilizes the majority spin electrons, but also increasingly localizes the minority spin Cr $3d_{z^2}$ electrons

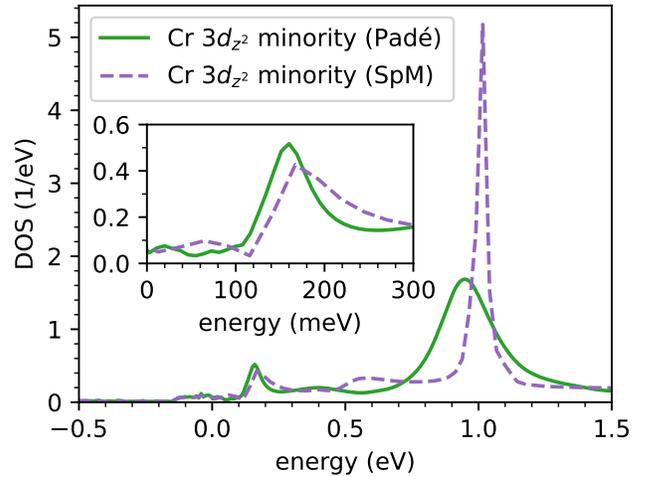

FIG. S13. DFT+DMFT spectral function for the minority-spin Cr $3d_{z^2}$ orbital of CrGeTe$_3$ at T = 100 K and P = 5 GPa in the ferromagnetic state. Results are shown for both Padé and sparse modeling (SpM) analytic continuation methods. The inset shows the low-energy feature, which persists independent of the continuation method.

in the vicinity of these holes, since they can lower their energy by hopping into a hole, where they are only sub-

(a)

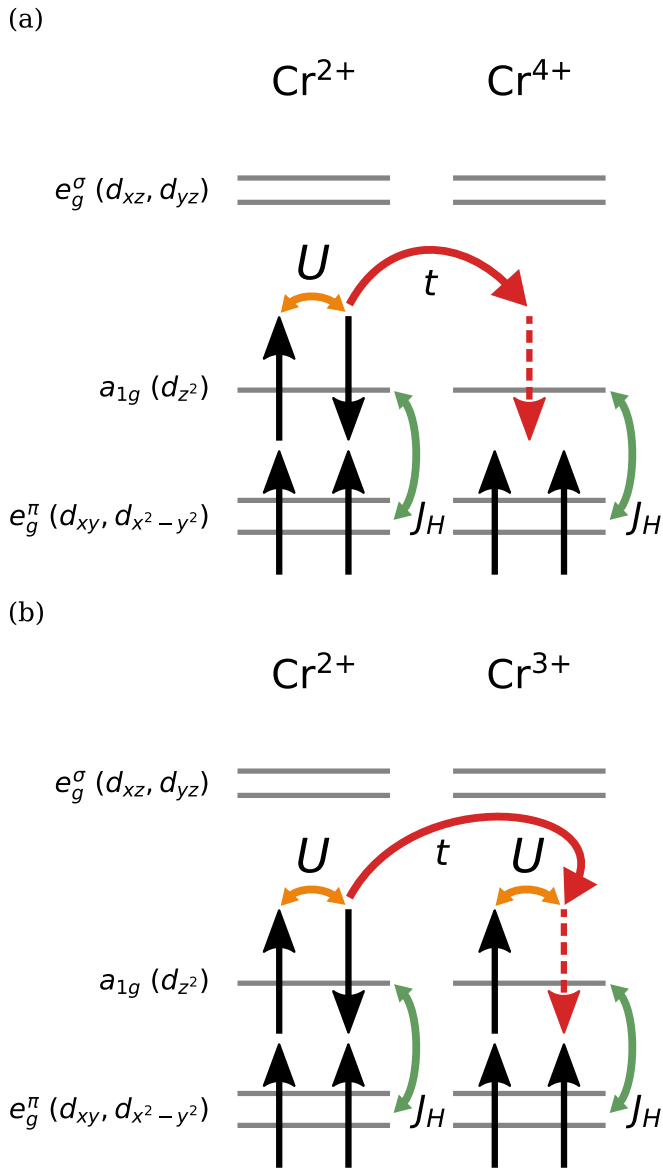

(b)

FIG. S14. Schematic depiction of exchange processes relevant for minority spin Cr $3d_{z^2}$ electrons. (a) Hopping of a minority spin electron into a majority spin $d_{z^2}$ hole. (b) Hopping of a minority spin electron onto an occupied majority spin $d_{z^2}$ site.

ject to Hund's rule coupling (see Fig. S14(a)), but avoid the Coulomb repulsion of a doubly occupied site (see Fig. S14(b)). Since the energy cost of spin misalignment with respect to Hund's coupling $J_H$ (see Fig. S14(a)) is roughly equal to the distance of both peaks in the spectral function (at $\pm 200$ meV), it seems possible that Hund's coupling is responsible for these low-energy features of the electronic self-energy (see Fig. S11(b)).

## VI. EXPERIMENTAL ESTIMATE OF MASS ENHANCEMENT

We can furthermore trace the strength of electronic correlations based on the experimental optical conductivity spectrum. The extraction of the electronic correlation strength from optical conductivity spectra is an established procedure applied to various quantum materials [36–38]. The optical conductivity spectrum of CrGeTe$_3$ in the high-pressure (either paramagnetic or ferromagnetic) metallic phase contains a MIR absorption band, which, according to our theoretical calculations, is attributed to electronic correlation effects. The spectral weights of the MIR band and the Drude term can serve as a measure for the electronic correlation strength, as explained below. This gives us the unique possiblity to trace the correlation strength as a function of pressure in a 2D vdW material.

The Drude spectral weight $\omega_p^2$ serves as an estimate of the optical kinetic energy $K_{opt}$ of the quasiparticles, and its reduction as compared to its value $K_{band}$ obtained from non-interacting band theory calculations is a measure of the electronic correlation strength [36–38]. The ratio $K_{opt}/K_{band}$ can be estimated from the experimental plasma frequency $\omega_p$ of the Drude term and the oscillator strength $\Omega_{MIR}$ of the MIR band according to [37]

$$ratio_{corr} = \frac{K_{opt}}{K_{band}} \approx \frac{\omega_p^2}{\omega_p^2 + \Omega_{MIR}^2} \quad . \tag{7}$$

The value of $ratio_{corr}$ ranges between 0 (Mott insulator) and 1 (uncorrelated metal). It corresponds to the quasiparticle weight calculated in section IV.

In case of CrGeTe$_3$, the so-obtained value of $ratio_{corr}$ as a function of pressure at 25 K is shown in Fig. 1(d) in the main text. At $P_c$, $ratio_{corr}$ rises sharply and saturates at the value 0.7 above 4.6 GPa Accordingly, CrGeTe$_3$ in its metallic phase is moderately correlated, similar to the square-net nodal-line semimetal ZrSiSe and slightly less correlated than the ferromagnetic kagome metal Co$_3$Sn$_2$S$_2$ [36]. In comparison, for strongly correlated metals such as cuprates and the vanadium oxide V$_2$O$_3$ a value $ratio_{corr} \sim 0.2$ would be expected [38] (see also Table S1 for a comparison with other vdW materials).

## VII. ESTIMATE OF THE CURIE TEMPERATURE FROM OPTICAL DATA

When entering the ferromagnetic state during cooling, significant changes occur in the high-energy reflectivity spectrum R$_{s-d}$, as illustrated in Figs. S15(a) and (c) for $P$=1.7 GPa and $P$=5.8 GPa, respectively. These temperature-induced changes appear even clearer in the first derivative of the reflectivity with respect to frequency $dR_{s-d}/d\nu$ [see Figs. S15(b) and (d)]. For example, at 1.7 GPa a clear dip feature develops in $dR_{s-d}/d\nu$

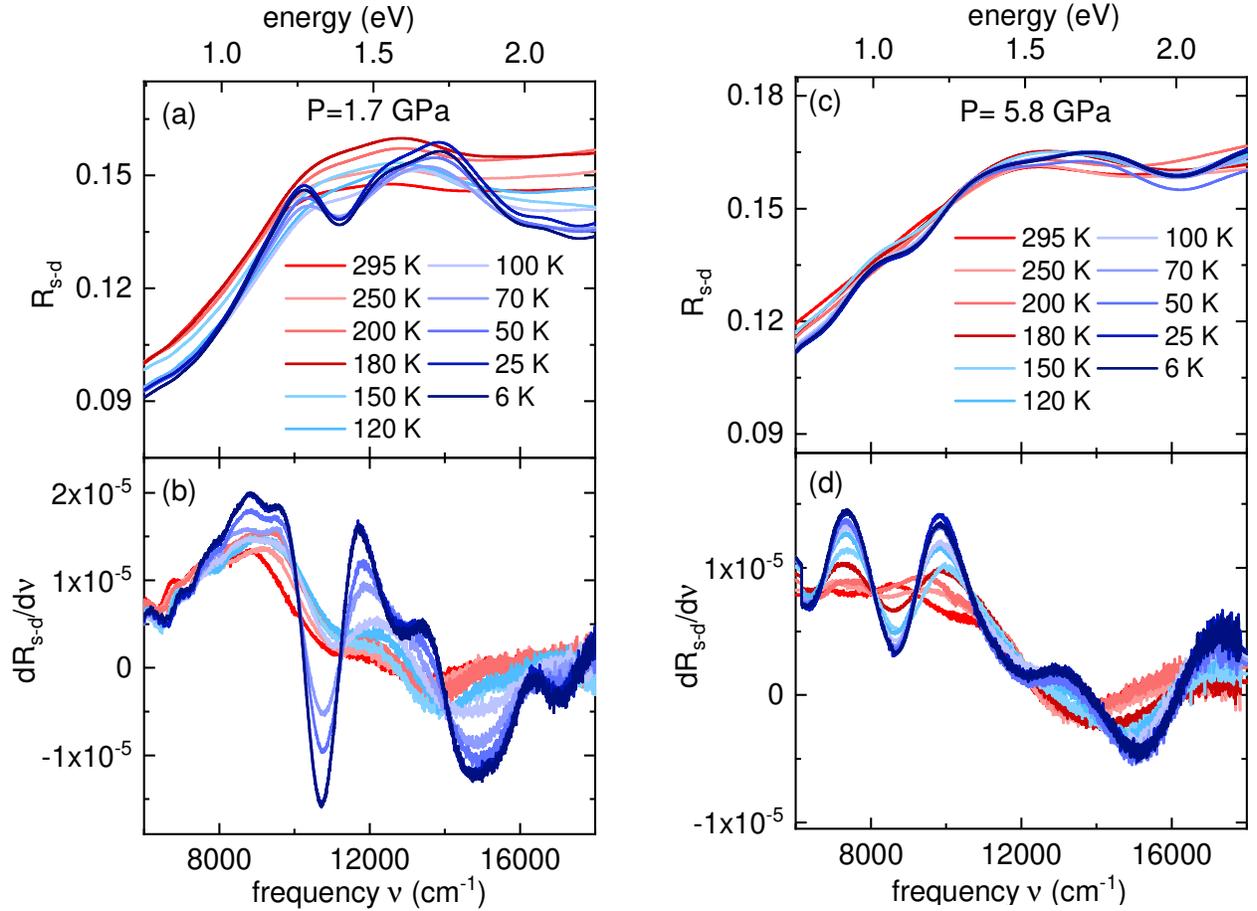

FIG. S15. High-energy reflectivity spectrum R$_{s-d}$ of CrGeTe$_3$ at 1.7 GPa (a) and at 5.8 GPa (c) together with the corresponding first derivative $dR_{s-d}/d\nu$ at 1.7 GPa (b) and at 5.8 GPa (d).

at $\sim$10.800 cm$^{-1}$ between 100 and 70 K [Fig. S15(b)]. Accordingly, magnetic order sets in below 100 K, and we can estimate the Curie temperature $T_C$=85 K $\pm$ 15 K at 1.7 GPa. Applying this criterion to all measured pressures, we obtained the pressure dependence of the magnetic ordering temperature $T_C$ as depicted in Fig. S16.

## VIII. DECOMPOSITIONS OF THE $\sigma_1$ SPECTRA AS A FUNCTION OF PRESSURE AND TEMPERATURE

The decompositions of the $\sigma_1$ spectra as a function of pressure and temperature are shown in Figs. S17, S18, S19 and S20.

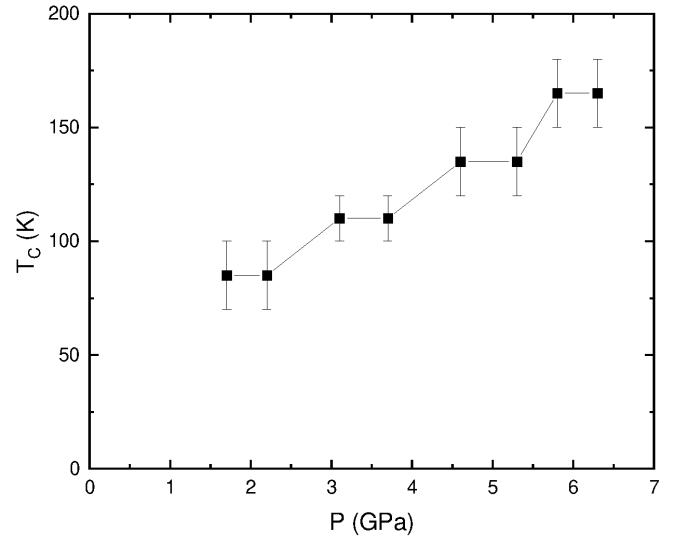

FIG. S16. Ferromagnetic ordering temperature $T_C$ of CrGeTe$_3$ as a function of pressure as obtained from the optical data.

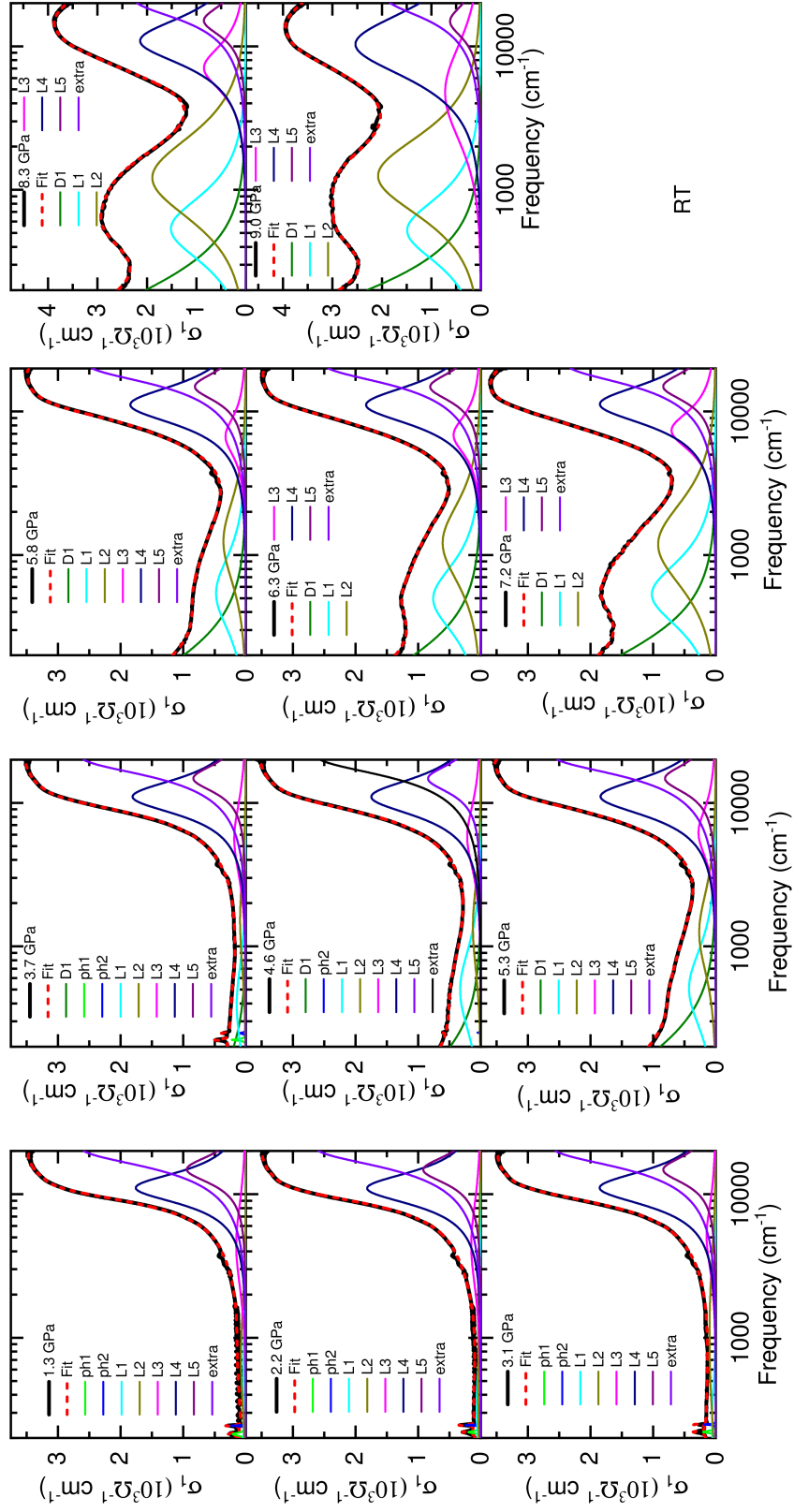

FIG. S17. Decompositions of the $\sigma_1$ spectra as a function of pressure at room temperature.

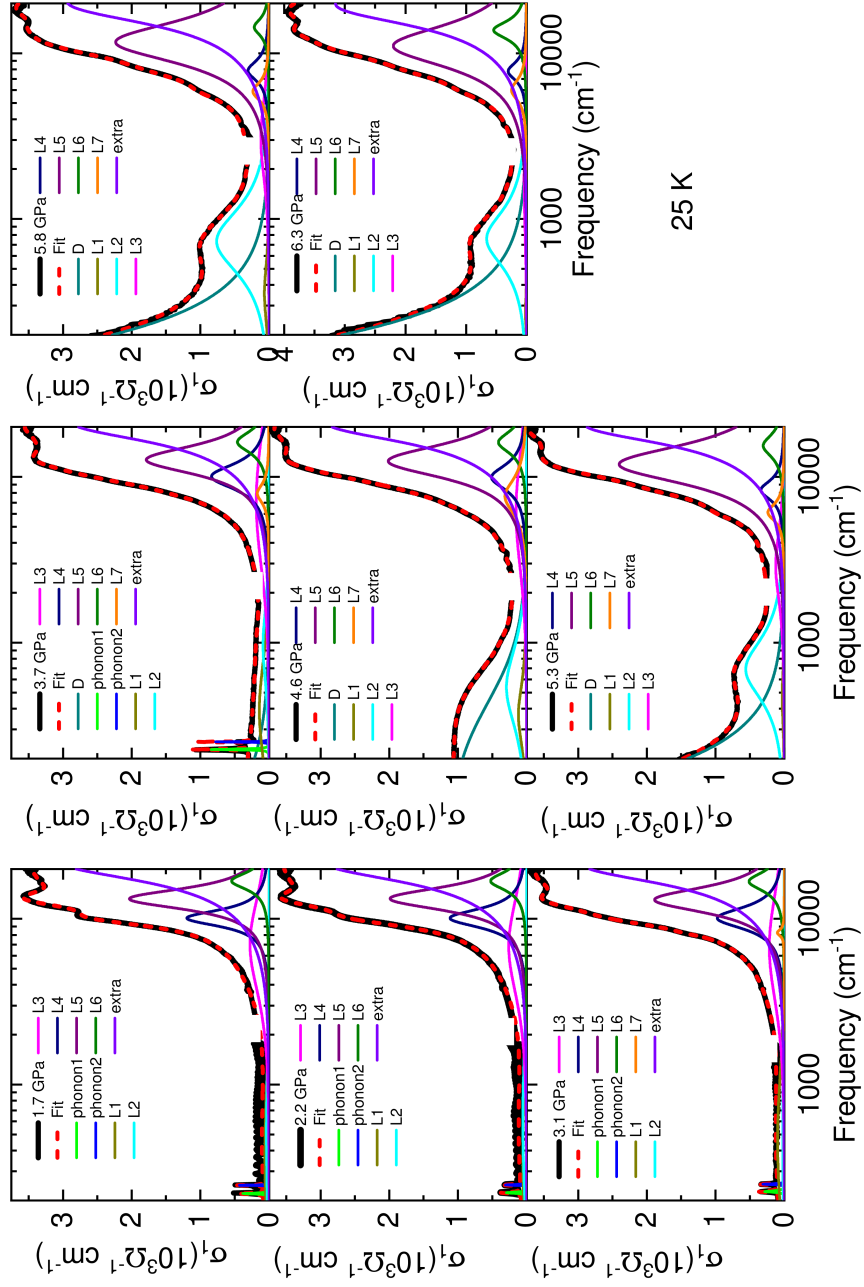

FIG. S18. Decompositions of the $\sigma_1$ spectra as a function of pressure at 25 K.

12

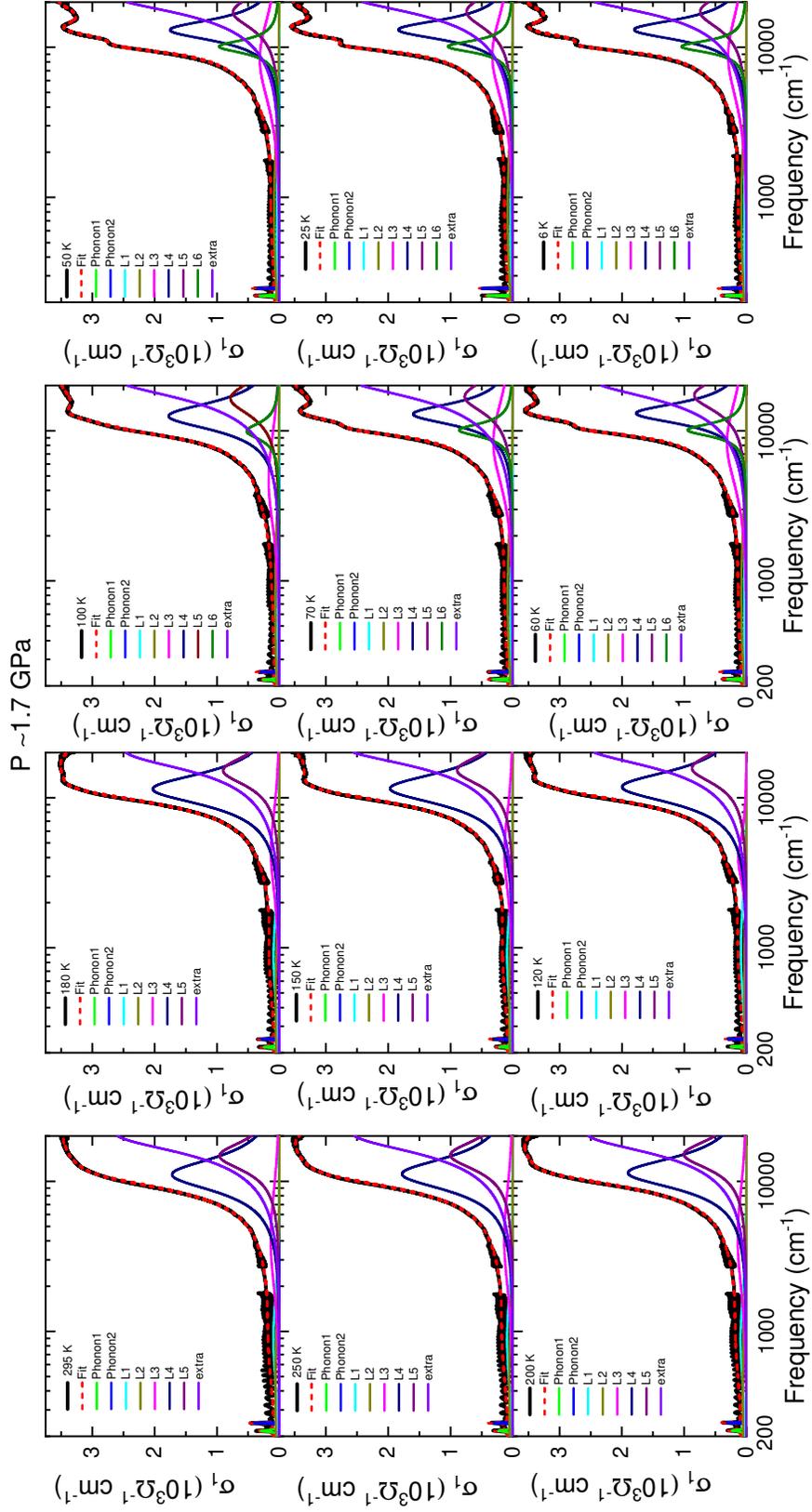

FIG. S19. Decompositions of the $\sigma_1$ spectra as a function of temperature at ~1.7 GPa.

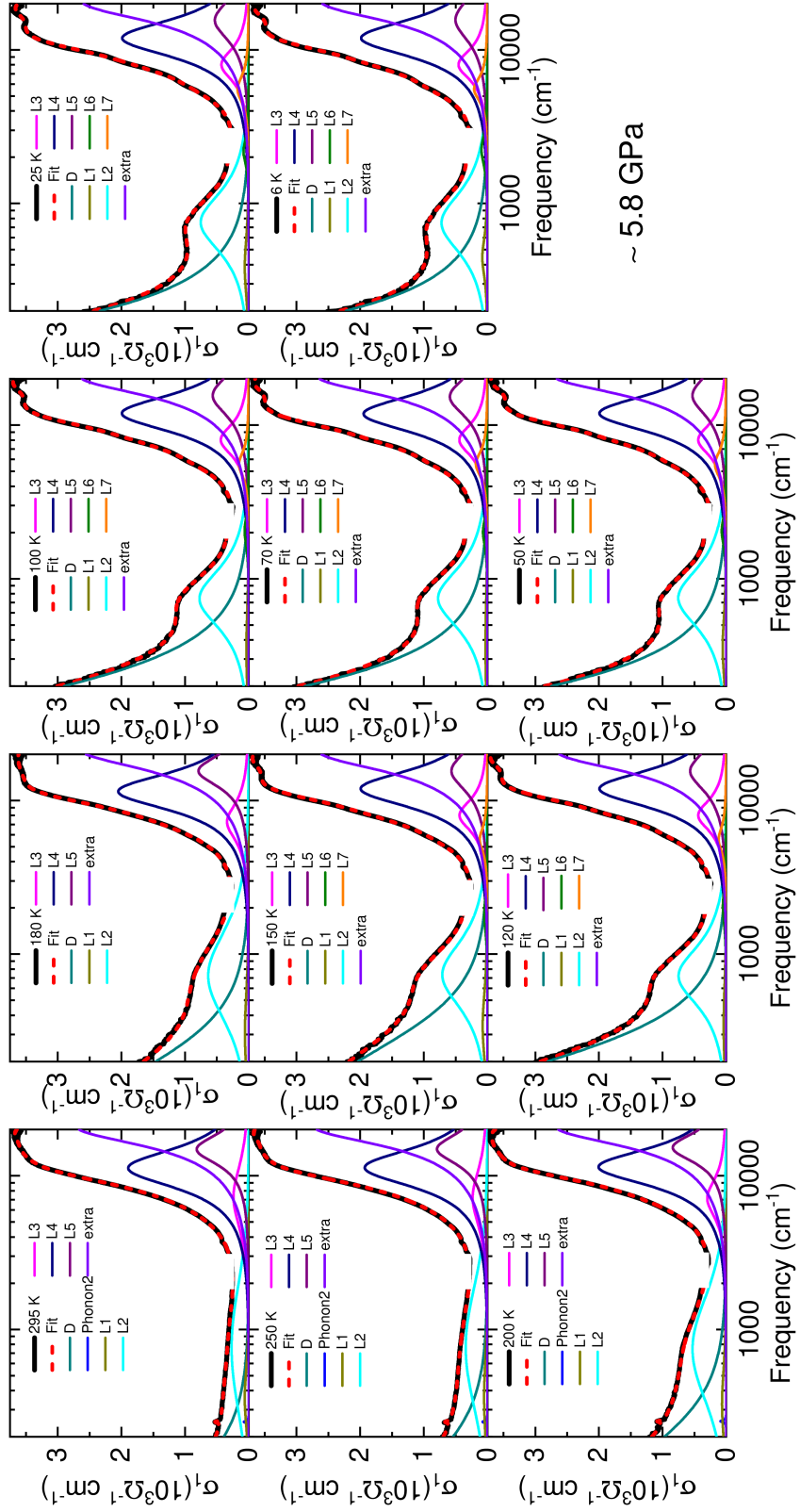

FIG. S20. Decompositions of the $\sigma_1$ spectra as a function of temperature at ~5.8 GPa.



\end{document}